\documentclass[preprint,12pt]{elsarticle}

\usepackage{epsf,epsfig,float,amssymb,latexsym}
\usepackage{rotating}

\usepackage{graphicx}

\usepackage{amssymb}

\newlength{\MySep}
\newcommand{\beq}{\begin{equation}}
\newcommand{\eeq}[1]{\label{#1}\end{equation}}
\newcommand{\beqa}{\begin{eqnarray}}
\newcommand{\eeqa}[1]{\label{#1}\end{eqnarray}}
\newcommand{\eeqan}{\end{eqnarray}}
\newcommand{\CR}{\nonumber \\ }

\newcommand {\MySepRule}   {\rule{0em}{1.425em}}

\newcommand {\BeNul}       {b_{0}}
\newcommand {\BeDe}        {b_{\rm D}}
\newcommand {\BeFe}        {b_{\rm F}}
\newcommand {\DeNul}       {d_{0}}
\newcommand {\DeDe}        {d_{\rm D}}
\newcommand {\DeFe}        {d_{\rm F}}
\newcommand {\DeI}         {d_{1}}
\newcommand {\DeII}        {d_{2}}
\newcommand {\DxD}         {D^{2}}
\newcommand {\DxF}         {DF}
\newcommand {\FxF}         {F^{2}}

\makeatletter
\def\gsim{\compoundrel>\over\sim}
\def\lsim{\compoundrel<\over\sim}
\def\compoundrel#1\over#2{\mathpalette\compoundreL{{#1}\over{#2}}}
\def\compoundreL#1#2{\compoundREL#1#2}
\def\compoundREL#1#2\over#3{\mathrel
      {\vcenter{\hbox{$\m@th\buildrel{#1#2}\over{#1#3}$}}}}
\makeatother

\journal{Nuclear Physics A}

\begin{document}

\begin{frontmatter}



\title{Chirally motivated separable potential model for $\eta N$ amplitudes}

\author[UJF]{A.~Ciepl\'{y}}
\author[UTEF]{J.~Smejkal}
\address[UJF]{Nuclear Physics Institute, 250 68 \v{R}e\v{z}, Czech Republic}
\address[UTEF]{Institute of Experimental and Applied Physics, Czech Technical University in Prague,\\
Horsk\'{a} 3a/22, 128~00~Praha~2, Czech Republic}

\begin{abstract}
We analyze the $\eta N$ interaction using a coupled channel separable potential model 
that implements the chiral symmetry. The model predicts an $\eta N$ stattering length 
$\Re a_{\eta N} \approx 0.7$ fm and in-medium subthreshold attraction most likely sufficient 
to generate $\eta$-nuclear bound states. The energy dependence of the $\eta N$ amplitude 
and pole content of the model are discussed. An idea of the same origin of the baryon 
resonances $N^{\star}(1535)$ and $N^{\star}(1650)$ is presented.
\end{abstract}


\end{frontmatter}


\section{Introduction}
\label{sec:intro}

A modern treatment of meson--baryon interactions at low energies is based on chiral perturbation theory ($\chi$PT)
that implements the QCD symmetries in its nonperturbative regime. The idea can be traced back to the late 
1970's when Weinberg came up with his phenomenological Lagrangians \cite{1979Wei}, the effective field theories 
which have a virtue in providing us with a systematic way to calculate the perturbative corrections. 
Being of this nature the $\chi$PT has proven itself as an effective theory that utilizes an expansion 
in powers of small momenta and light quark masses. Naturally, the theory is expected to work well in the SU(2) 
sector due to smallness of the {\it up} and {\it down} quark masses. Interestingly, an extension to the SU(3) 
sector \cite{1985GL} has lead to a very successful description of $\bar{K}N$ interactions despite much larger mass 
of the strange quark and presence of the $\Lambda(1405)$ resonance just below the $\bar{K}N$ threshold 
\cite{1995KSWnpa}, \cite{1998OR}. There, the $\chi$PT is helped by a classical resummation technique, 
the Lippmann-Schwinger equation, which enables to sum up the major part of the perturbation series. 
As a result the higher order corrections are accounted for in a situation when the standard perturbation 
approach does not converge. 

In the pioniering works \cite{1995KSWnpa}, \cite{1995KSWplb} the authors introduced local and separable potentials 
that match the chiral meson-baryon amplitudes up to a given perturbation order. The range parameters that appear 
in such approach do not only help to regularize the intermediate state Green function integral 
but they also provide a natural off-shell extrapolation of the amplitudes. This feature makes 
the model suitable for in-medium applications, particularly for a construction of meson-nuclear 
optical potential. We have followed this path in our recent works \cite{2010CS}, \cite{2012CS} 
to establish a microscopical understanding of antikaon iteractions with nuclei.

Another way to manage the meson-baryon interactions reflects the unitarity of the scattering 
S-matrix and is based on a dispersion relation for the inverse of the T-matrix 
or on the N/D method \cite{1998OR}, \cite{2001OM}, \cite{2003JOORM}. There, the intermediate 
state Green function is regularized by standard quantum field techniques, most commonly 
by dimensional regularization. Both approaches lead to equivalent description (compare 
e.g.~the two most recent analysis \cite{2012CS} and \cite{2012IHW}, both including the precise 
kaonic hydrogen data from the SIDDHARTA collaboration \cite{2011SIDD}) of the energy dependence 
of the meson-baryon amplitudes with the subtraction constants introduced in dimensional 
regularization non-trivially related to the range parameters employed in the potential model. 

The chiral coupled channels approaches to $\bar{K}N$ interaction have brought quite new insights 
on the dynamics of the meson baryon interactions in the free space as well as in the nuclear medium. 
One of the most interesting results appears to be the two-pole character of the $\Lambda(1405)$ 
resonance \cite{2003JOORM} which is formed dynamically due to a strong inter-channel coupling 
in the $\pi \Sigma$--$\bar{K}N$ system \cite{2008HW}. The dynamics of the $\Lambda(1405)$ 
then leads to a strong energy dependence of the $\bar{K}N$ amplitude not only in the free space 
but in nuclear matter as well \cite{2011CFGGM}, \cite{2012CS}. It has been well known for some years 
that the $\bar{K}$-nuclear optical potentials constructed from the chirally motivated $\bar{K}N$ 
amplitudes taken at the threshold energy are quite shallow \cite{2000RO}, \cite{2001CFGM} 
while a phenomenological analysis of kaonic atoms revealed very attractive deep $\bar{K}$-nuclear 
potentials \cite{1994FGB}. It was demonstrated in Ref.~\cite{2011CFGGM} that these two conflicting scenarios 
can be reconciled by considering properly the subthreshold energy dependence of the $\bar{K}N$ 
amplitude. Indeed, it appears that the relatively weak $\bar{K}N$ attraction becomes much stronger 
at energies about 30-50 MeV below the $\bar{K}N$ threshold that are probed by antikaons in nuclear 
matter, although even more attraction is needed phenomenologically. This strong attraction 
is also in agreement with predictions by relativistic mean field theories that implement 
the antikaon field \cite{2006MFG}. It comes without saying that a good 
understanding of (anti)kaon interaction with nuclei would not be possible without realistic models 
to describe the subthreshold energy dependence of the $\bar{K}N$ amplitudes in the free space 
as well as in the nuclear medium. The potential we used to study the $\bar{K}N$ interactions 
perfectly suits this role.

The situation is quite similar in case of the $\eta N$ interaction whose energy dependence around 
the threshold is also strongly affected by a resonance, this time the $N^{\star}(1535)$ one. Thus, it seems 
natural to utilize the techniques developed for the $\bar{K}N$ system and apply them to the $\eta N$ one. 
This is exactly what represents the content of our current work. In close resemblance to our $\bar{K}N$ 
model \cite{2010CS}, \cite{2012CS}, here we study the $\eta N$ system within a chirally motivated coupled channels framework.
In fact, the same approach was already used in Refs.~\cite{1995KSWplb}, \cite{1997KWW} and \cite{1997WW} 
though the authors of those works restricted their analysis to a relatively narrow energy interval 
above the $\eta N$ threshold. We are also in a more favourable position thanks to new precise data 
on the $\pi N \longrightarrow \eta N$ reaction \cite{2005CBC} as well as due to an existence of 
a more advanced database of the $\pi N$ partial waves \cite{SAID}. We note that the $\eta N$ 
interaction was also recently studied by other authors who used the unitary coupled channel approach 
based on solving the Bethe-Salpeter equation \cite{2001NR}, \cite{2002IOV} and \cite{2012MBM}. Excepting Ref.~\cite{2001NR} 
these models predict a moderately attractive $\eta N$ interaction with a scattering length 
$\Re a_{\eta N} \approx 0.3$ fm while a considerably larger attraction was obtained in $K$-matrix 
fits to $\pi N$  and $\gamma N$ reaction data with $\Re a_{\eta N} \approx 1$ fm \cite{2005GW}. 
As we will show, our model yields $\Re a_{\eta N} \approx 0.7$ fm, a value approximately in-between 
those two estimates and in agreement with Refs.~\cite{1995KSWplb} and \cite{2001NR}. Naturally, 
the strength of the $\eta N$ attraction at the threshold and at energies below the $\eta N$ threshold 
are relevant for a possible existence of $\eta$-nuclear bound states \cite{2013FGM}. The authors 
of Ref.~\cite{2013FGM} found that an attraction related to $\Re a_{\eta N} \approx 0.3$ fm might 
be just sufficient to generate $\eta$-nuclear bound states in the $^{24}$Mg isotope (and in heavier targets) 
while a stronger attraction is required for binding the $\eta$ in lighter nuclear systems.

The paper is organized as follows. After a brief introduction of the model in Section \ref{sec:model} 
we detail the procedure to fit the model parameters and our treatment of the experimental data 
in Section \ref{sec:fits}. In Section \ref{sec:results} we make comparisons with approaches 
by other authors and present the results of our own fits. The section is also continued with 
a discussion of the energy dependence of the $\eta N$ amplitudes in the free space. 
Finally, in Section \ref{sec:poles} we examine the pole content of the model and its relevance 
to the $N^{\star}(1535)$ and $N^{\star}(1650)$ resonances. The paper is closed with a short Summary.

\section{Coupled channel chiral model}
\label{sec:model}

The meson-baryon effective potential employed in our work is given in a separable form 
\beq
V_{ij}(k,k';\sqrt{s}) =  g_{i}(k^{2}) \: v_{ij}(\sqrt{s}) \: g_{j}(k'^{2})
\eeq{eq:Vsep}
with the off-shell form factors chosen in the Yamaguchi form,
\beq
g_{j}(k)=1/[1+(k/ \alpha_{j})^2] \;\;\; .
\eeq{eq:FF}
Here the indexes $i$ and $j$ run over the coupled meson-baryon channels and we number 
them in order of their threshold energies. Further, $k$ ($k'$) represents the CMS 
meson-baryon momenta in the initial (final) state, $\sqrt{s}$ 
stands for the two-body CMS energy, and the inverse ranges $\alpha_j$ characterize 
the interaction range of the specific meson-baryon states. The central piece 
of the chirally motivated potential matrix reads
\beq
v_{ij}(\sqrt{s}) = -\frac{C_{ij}(\sqrt{s})}{4\pi f_{i}f_{j}}\: \sqrt{\frac{M_i M_j}{s}}           
\eeq{eq:vij}
where $f_{j}$ is the meson decay constant and $M_{j}$ denotes the baryon mass 
in the $j$-th channel. Following the approach of other authors \cite{2001NR}, \cite{2002IOV}, 
\cite{2012IHW} we (in principle) allow for different values of $f_j$ that relate 
to the specific meson appearing in the $j$-th channel. 

The coupling matrix $C_{ij}$ is energy dependent and its form is determined by the chiral SU(3) 
symmetry. At the leading order Tomozawa-Weinberg (TW) interaction it is given as
\beq
C_{ij}(\sqrt{s}) = - C_{i j}^{\rm (TW)} (2\sqrt{s} -M_{i} -M_{j})/4
\eeq{eq:Cij}
with the TW couplings $C_{i j}^{\rm (TW)}$ standing for the standard SU(3) 
Clebsch-Gordan coefficients. In our work we also take into account the next-to-leading order (NLO) 
contributions to the $C_{ij}$ matrix that contribute at the $\mathcal{O}(q^2)$ order 
of the external meson momenta. The relation to a chiral expansion in terms 
of small meson momenta and quark masses is guaranteed 
by matching the pseudopotentials Eq.~(\ref{eq:vij}) to the meson-baryon amplitudes 
evaluated within the $\chi$PT to a given order, see Refs.~\cite{1995KSWnpa}, 
\cite{2010CS} for details. In Born approximation and for the momenta on the energy shell 
the meson-baryon amplitudes are equivalent with the pseudopotentials defined 
by Eq.~(\ref{eq:vij}). Thus, a positive sign of the pseudoponetial $v_{jj}$ (or a negative 
sign of the pertinent diagonal coupling $C_{jj}$) corresponds to an attractive interaction 
in the $j$-th channel. We also note in passing that a slightly different energy 
dependence of the TW term was used in some of our earlier works based on the heavy baryon 
formulation of the underlying chiral Lagrangian. The form adopted here and defined 
by Eq.~(\ref{eq:Cij}) reflects the fact that physical baryon masses do differ from 
a common baryon mass in the chiral limit. 

\begin{figure}[h]
\label{fig:FD}
\caption{Visualization of the LO and NLO contributions to the meson-baryon inter-channel 
couplings. The initial or final state mesons and baryons are denoted by $\phi_{i/j}$ and $B_{i/j}$, 
respectively. The four graphs represent: a) $\mathcal{O}(q)$ contact TW term, 
b) direct s-term, c) crossed u-term, and d) $\mathcal{O}(q^2)$ contact terms.}
\centering
\includegraphics[trim=0cm 0cm 0cm 24.5cm, clip=true, width=\textwidth]{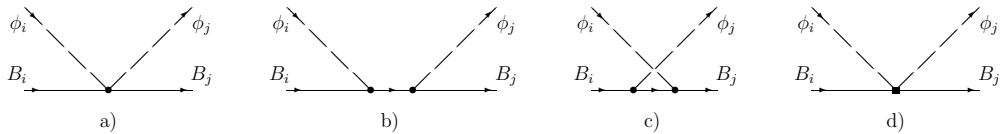}
\end{figure}

As we already stated the contributions included in our inter-channel couplings reflect the structure 
of the underlying chiral Lagrangian. Instead of writing up all the pieces that appear in  
the Lagrangian at the first and second chiral orders (for details see e.g.~Refs.~\cite{1995KSWnpa}, 
\cite{2006OVP}) we illustrate the contributions relevant for an $s$-wave scattering in a form 
of Feynman diagrams shown in Figure~\ref{fig:FD}. The first diagram visualizes the already 
mentioned current algebra TW term that plays a major role for the meson-baryon interactions 
at low energies and some analysis restrict themselves to only this interaction. 
The direct s-term and crossed u-term represent Born amplitudes that also originate from the 
first order chiral Lagrangian. When expanded in terms of external meson momenta the first 
three diagrams include not only the $\mathcal{O}(q)$ order but give also rise to relativistic 
corrections of the $\mathcal{O}(q^2)$ and higher orders. In addition we also incorporate 
the NLO terms from the second order chiral Lagrangian that are represented by the contact 
diagram shown as the last one in Fig.~\ref{fig:FD}. These $\mathcal{O}(q^2)$ contributions 
can be viewed as NLO corrections to the TW term. The exact form of all contributions related 
to the depicted diagrams, that we use to build the coupling matrix $C_{ij}$, as well as 
the involved parameters (low energy constants) of the model are specified in the Appendix.

The separable potentials defined by Eqs.~(\ref{eq:Vsep}) and (\ref{eq:vij}) are then used 
in the Lippmann-Schwinger equation that allows us to solve exactly the loop series and obtain 
the meson-baryon amplitudes in a separable form as well,
\beq
F_{ij}(k,k';\sqrt{s}) = g_{i}(k^{2}) \: f_{ij}(\sqrt{s}) \: g_{j}(k'^{2}) \;\;\; .
\eeq{eq:Fsep}
The algebraical solution for the reduced (stripped off the form factors) amplitudes then 
reads
\beq
f_{ij}(\sqrt{s}) = \left[ (1 - v \cdot G(\sqrt{s}))^{-1} \cdot v \right]_{ij}
\eeq{eq:fij}
with an intermediate state meson-baryon Green function
\beq
G_{n}(\sqrt{s}) = -4\pi \: \int
\frac{d^{3}p}{(2\pi)^{3}}\frac{g_{n}^{2}(p^{2})}{k_{n}^{2}-p^{2} +{\rm i}0} 
 = \frac{(\alpha_n + {\rm i}k_n)^{2}}{2\alpha_{n}}\:[g_{n}(k_{n})]^{2} 
\;\;\; .
\eeq{eq:Gn}
Here we assume that the interaction occurs in a free space which also allows 
us to perform the analytical integration in Eq.~(\ref{eq:Gn}). The impact 
of nuclear medium can be implemented easily in the model with the transition amplitudes 
(\ref{eq:fij}) acquiring a density dependence \cite{2011CFGGM}, \cite{2012CS}. 
In such a case, the Green function (\ref{eq:Gn}) 
would also become density dependent with the free space propagator modified due to Pauli blocking 
\cite{1996WKW} and by the meson and baryon selfenergies \cite{1998Lut}. 

The dynamics of the system is defined by the involved meson-baryon channels and 
by the inter-channel couplings $C_{ij}$ derived from the $\chi$PT. In the present work 
we apply the model to study the $\eta N$ interactions and include all the channels 
that represent the interactions of the lightest meson octet with the lightest baryon 
octet. For practical reasons we will work in the isospin basis of states and treat 
separately the channels with isospins $I=1/2$ and $I=3/2$. Thus, we have four $I=1/2$ 
coupled channels ($\pi N$, $\eta N$, $K \Lambda$ and $K \Sigma$) and two $I=3/2$ channels 
($\pi N$ and $K \Sigma$). Although we aim exclusively at the $\eta N$ interaction 
that is restricted to the $I=1/2$ isospin the $I=3/2$ channels are involved in fits 
of the free parameters of the model. In both sectors the channels are ordered according 
to the channel threshold energies. We assign the channel indexes $j$ in the just specified order, 
reserving the first four indexes $j=1,...,4$ to the $I=1/2$ channels and indexes $j=5,6$ 
to the $I=3/2$ channels. This allows for a compact analysis with only one coupling matrix 
$C_{ij}$ that is specified completely in the Appendix. Of course, one could also work with 
six physical charged channels ($\pi^{0}n$, $\pi^{-}p$, $\eta n$, $K^{0} \Lambda$, 
$K^{0} \Sigma^{0}$, $K^{+} \Sigma^{-}$) which would allow for a proper treatment 
of threshold effects. However, as we intend to match the experimental data on the $\pi N$ 
amplitudes that are presented as partial waves with a well defined isospin and the $\eta N$ 
state itself has also a well defined isospin, we feel that doing the analysis in the isospin 
states basis is more appropriate. If a need arises the model can easily be adapted to work 
with the physical meson-baryon states too.

\section{Fits to experimental data}
\label{sec:fits}

The chirally motivated potential model described in the previous section has quite many 
parameters that have to be either fixed by making reasonable assumptions or fitted to 
the available experimental data. First of all there are the meson weak decay constants $f_{j}$ 
which we fix at their physical values $f_{\pi} = 92.4$ MeV, $f_{K} = 113.0$ MeV and 
$f_{\eta} = 120.1$ MeV \cite{1985GL}, depending on the meson in the respective channel. The strengths  
of the direct and crossed Born terms are characterized by combinations of the axial vector 
coupling constants $D=0.80$ and $F=0.46$ with the commonly accepted values taken from 
an analysis of semileptonic hyperon decays \cite{1999Rat}. The $b$-couplings of 
the NLO contact terms and the common baryon mass $M_0$ were fixed in Refs.~\cite{2007CS} 
and \cite{2010CS} to satisfy the Gell-Mann formulas for baryon mass splittings and 
to guarantee a chosen value of the pion-nucleon sigma term $\sigma_{\pi N}$. 
Further, there are four independent $d$-couplings that determine the strengths of the double 
derivative contributions to the NLO contact terms, to the last diagram of Fig.~\ref{fig:FD}. 
These four parameters and the inverse ranges $\alpha_{j}$ that characterize the Yamaguchi 
form factors, Eq.(\ref{eq:FF}), are left free to be fitted to the experimental data.
In summary, we are left with eight free parameters, four $d$-couplings and four inverse 
ranges $\alpha_{j}$. 

To perform the fits we use the MINUIT routine from CERNLIB to minimize the $\chi^{2}$ per degree 
of freedom defined as
\beq
\chi^{2}/dof = \frac{\sum_{i}N_{i}}{N_{obs}(\sum_{i}N_{i}-N_{par})} \sum_{i}\frac{\chi^{2}_{i}}{N_{i}}
\eeq{eq:chi2}
where $N_{par}$ is the number of fitted parameters, $N_{obs}$ is a number of observables, $N_{i}$ 
is the number of data points for an $i$-th observable, and $\chi^{2}_{i}$ stands for the total $\chi^{2}$ 
computed for the observable. Eq.~(\ref{eq:chi2}) guaranties an equal weight of all fitted data.

There is a plenty of experimental data on low energy $\pi N$ scattering and reactions. 
The model parameters are fitted to
\begin{itemize}
\item $\pi N \longrightarrow \pi N$ amplitudes for the $S_{11}$ and $S_{31}$ partial waves taken 
from the SAID database \cite{SAID}
\item selected $\pi^{-} p \rightarrow \eta n$ reaction total cross section data
\end{itemize}

Primarily, we make use of a comprehensive analysis of the available $\pi N$ data 
and match our $s$-wave $\pi N$ amplitudes in the $S_{11}$ and $S_{31}$ partial waves to those 
from the SAID database \cite{SAID}. Taking into account the normalization of the SAID amplitudes 
this is done by multiplying our amplitude $F_{\pi N, \pi N}$, Eq.~(\ref{eq:Fsep}), 
by the magnitude of the CMS momentum in the initial channel,
\beq
k_{\pi N} \: F_{\pi N, \pi N}(S_{11}/S_{31}) = F_{\rm SAID}(S_{11}/S_{31}) \;\;\; ,
\eeq{}
where we refer explicitly to a given partial wave. Since we are interested mainly 
in the low energy region close to the $\eta N$ threshold and up to about $50-80$ MeV subthreshold 
we restrict ourselves to meson-baryon CMS energies below $1600$ MeV. In the energy interval 
from the $\pi N$ threshold up to $1600$ MeV there are 30 single energy data points for each of 
$\Re F(S_{11})$, $\Im F(S_{11})$, $\Re F(S_{31})$, $\Im F(S_{31})$. Considering the precision of the SAID 
analysis and comparing it with a previous one by the Karlsruhe-Helsinki group \cite{1986KH}, we follow 
the approach of Ref.~\cite{2012MBM} and assume a semiuniform absolute variation of the SAID 
amplitudes. This variation is set to $0.005$ fm for energies below 1228 MeV, and to $0.03$ fm 
for energies above 1228 MeV. Finally, it is apparent in the SAID analysis that the experimental 
$S_{31}$ amplitude is srongly affected by the $\Delta(1620)$ resonance at the higher end of our 
energy interval. Since this resonance does not have a prominent dynamically generated component 
and is completely missing in our model we exclude the affected $S_{31}$ amplitudes from our fits.
A comparison of the $S_{31}$ amplitudes generated by our model with those from the SAID analysis 
has shown that we can safely consider the SAID data in the $I=3/2$ sector up to $1450$ MeV. 
This means that for the $\Re F(S_{31})$ and $\Im F(S_{31})$ amplitudes we exclude from the fit 
the SAID data at 9 highest single energies.

In addition to the $\pi N$ amplitudes we also fit the $\pi^{-} p \rightarrow \eta n$ 
reaction cross sections. In our formalism, the total $s$-wave cross sections 
for a transition from $j$-th to $i$-th channel, are given by the standard formula, 
\beq
\sigma_{ij} = 4\pi\: \frac{k_{i}}{k_{j}} \: |F_{ij}|^{2}
\;\;\; .
\eeq{eq:xsect} 
In the isospin basis, the experimental $\eta n$ production cross section can be identified with 
\mbox{$2/3\, \sigma_{I=1/2}(\pi N \rightarrow \eta N)$} calculated within our coupled channels 
model. There are very precise experimental data on the $\eta n$ production from a recent 
measurement by the Crystall Ball collaboration \cite{2005Pra}. The precision of these 
data exceeds by far the much older data from previous experiments. For this reason we consider 
only the new data from Ref.~\cite{2005Pra} in the energy region above the $\eta N$ threshold 
and complement them with data from three other older measurements \cite{1969Bul}, \cite{1970Ric}, 
\cite{1975Fel} only at energies above $1525$ MeV, the highest energy treated in Ref.~\cite{2005Pra}. 
In accordance with our approach to the $\pi N$ amplitudes we also restrict ourselves to energies 
below $1600$ MeV. This means that a $p$-wave contribution to the fitted total $\eta n$ 
cross sections can be neglected. However, to match properly the experimental $\eta n$ 
production data we modify the calculated cross sections to account for a lack of the three-body 
$\pi \pi N$ channel in our model. This channel decreases the experimental inelasticity of 
the $\pi N$ amplitude reported in the SAID database. The total reaction cross section reads as
\beq
\sigma_{r} = \frac{\pi}{k_{\pi N}^{2}} (1 - \eta^{2})
\eeq{eq:sigr}
The SAID database provides a factor $(1 - \eta^{2}) = 0.917$ at the energy $\sqrt{s} = 1540$ MeV 
which gives a total reaction cross section of about $3.5$ mb, approximately 20\% larger than 
the maximum of the experimental $\pi^{-}p \rightarrow \eta n$ cross section. In our fit procedure 
we effectively compensate for the missing $\pi \pi N$ channel by enhancing the calculated $\eta N$ 
cross sections that represents in our model the only reaction channel which is open in the discussed 
energy region. Thus, the calculated $\eta N$ cross section $\sigma_{I=1/2}$ is matched 
to the experimental one by using a relation
\beq
\sigma(\pi^{-}p \rightarrow \eta n) = \frac{2}{3} \sigma_{I=1/2}(\pi N \rightarrow \eta N) /1.2
\eeq{eq:pipiN}
We note that an introduction of the factor of $1.2$ is in agreement with observations made 
in Ref.~\cite{2003GHH}. We also found that fits made without this factor lead to only slightly worse 
reproduction of the $\eta N$ production data while the effect on the overall quality and 
other characteristics of the fit is not significant.

\section{Results}
\label{sec:results}

\subsection{The KSW model}

\begin{figure}[h]    
\centering
\includegraphics[width=0.45\textwidth]{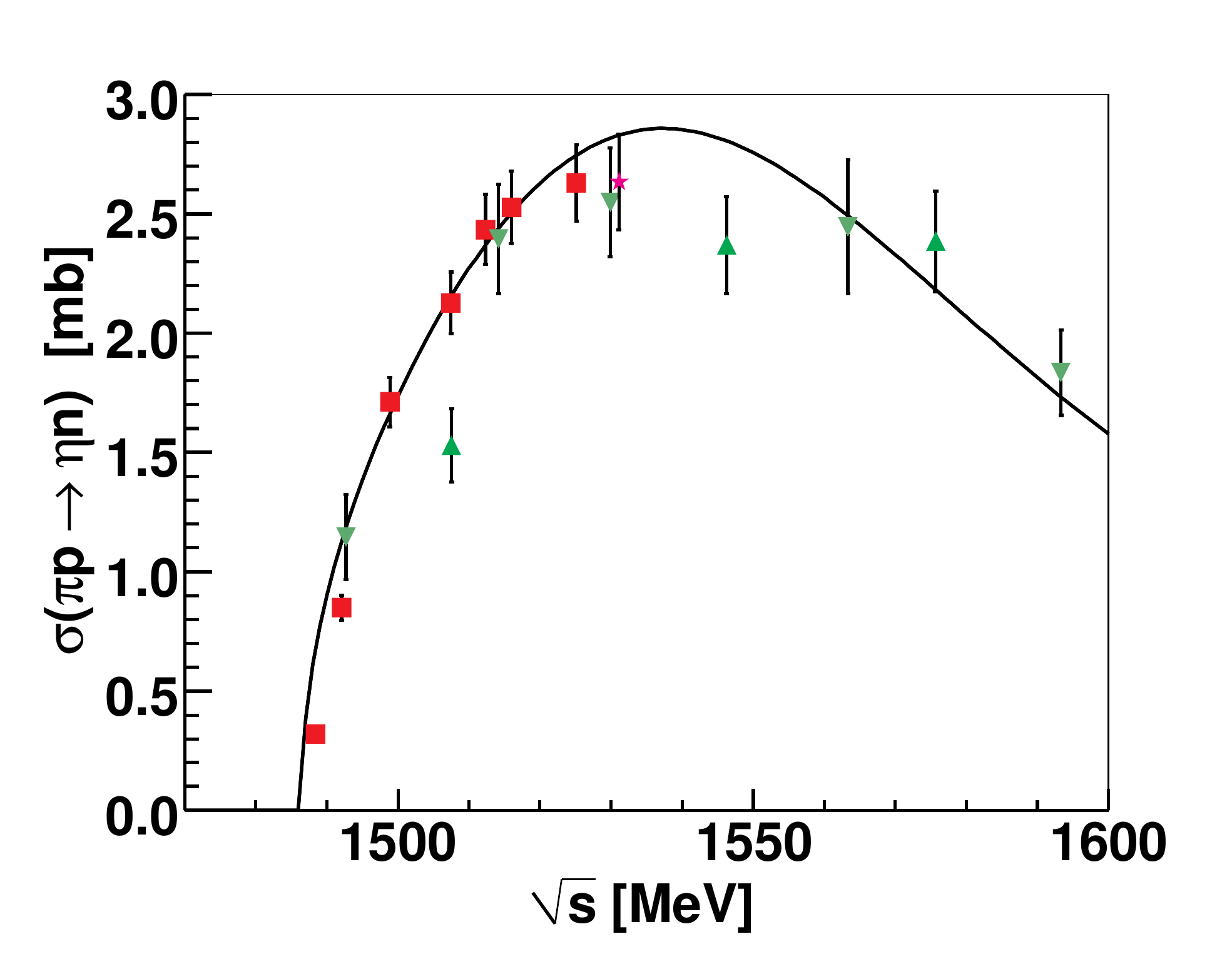}
\includegraphics[width=0.45\textwidth]{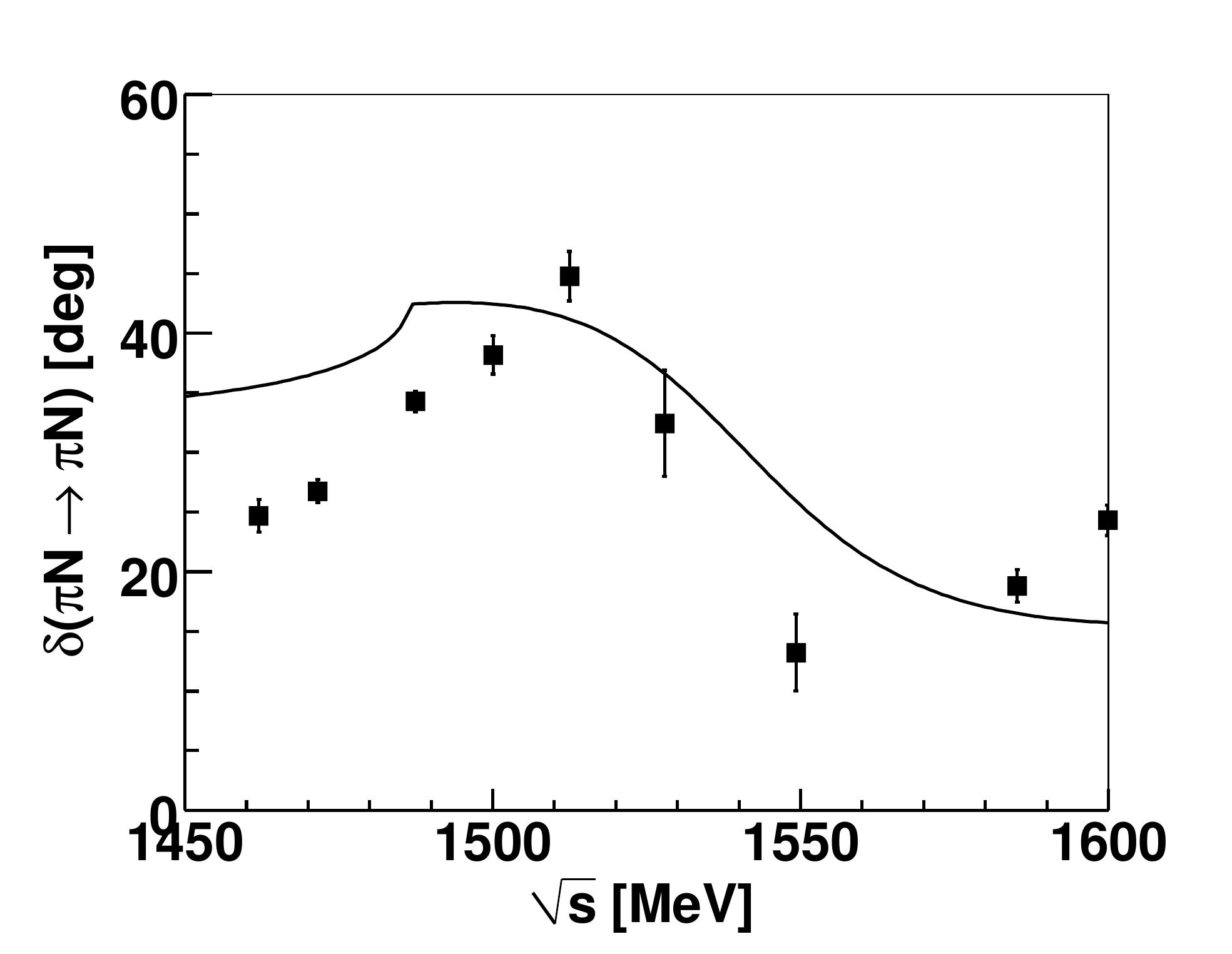}
\caption{Reproduction of the KSW results \cite{1995KSWplb}. Left panel: $\pi N \rightarrow \eta N$ cross section, 
right panel: $\pi N \rightarrow \pi N$ $S_{11}$ phase shift. The experimental data are taken from the current 
version of the SAID database \cite{SAID}, older data were shown in the original KSW paper.}
\label{fig:KSW}
\end{figure}

Before presenting the results of our fits to the experimental data we check the functionality 
of the model and the pertinent computer code by reproducing the old results by Kaiser, Siegel and Weise 
\cite{1995KSWplb}. They used essentially the same effective meson-baryon potentials with an alternate 
energy dependence of the TW term and omitted contributions from the Born diagrams, 
the direct and crossed terms shown in Figure \ref{fig:FD}. The two Born diagrams were included in a following 
work \cite{1997KWW} in which the authors got an equivalent description of the $\eta N$ data. 
However, a comparison of these later results with our current work is not so straightforward 
due to a different form of the off-shell form factors employed in Ref.~\cite{1997KWW}.
In what follows we will refer to the original model of Ref.~\cite{1995KSWplb} by the KSW tag 
and use it for a comparison with our own fits. Adopting the KSW potential form and parameters 
of the KSW model we were able to reproduce nicely their results. They are shown 
in Figure \ref{fig:KSW} which is to be compared with Figure 1 and with the left panel 
of Figure 2, both of them published in Ref.~\cite{1995KSWplb}. The experimental data 
on the $\eta N$ production are taken from Refs.~\cite{2005Pra}, \cite{1969Bul}, \cite{1970Ric} 
and \cite{1975Fel}. The first three data points marked by triangles in our Figure \ref{fig:KSW} 
were not included in our fits and are given in the figure to show the superiority of the recent 
experimental data over the older ones at energies closely above the $\eta N$ threshold. 
The experimental data on the $S_{11}$ phase shifts are taken directly from the SAID database \cite{SAID}.

Concerning a comparison of the KSW model with our own fits one should remember that the KSW 
model was designed to reproduce the experimental data only in a relatively narrow region 
of the $N^{\star}(1535)$ resonance, in an interval limited to the CMS energies 
$1450$ MeV $< \sqrt{s} < 1600$ MeV. We found that the $\pi N$ amplitudes generated by the model 
start to deviate quite significantly from those of the SAID analysis as soon as one goes 
below the $\eta N$ threshold. Thus, the KSW model is clearly not suitable for description 
of the $\pi N$ and $\eta N$ data in the whole energy region $m_{\pi}+M_{N} < \sqrt{s} < 1600$ MeV 
considered here.

\subsection{New fits and results}

The results of our fits are shown in Table~\ref{tab:fits} in comparison with the KSW model.
To have some variety and check a consistency of our model predictions we opted for three different 
choices of the pion-nucleon sigma term that fixes the $b_0$ parameter and the common baryon mass 
in the chiral limit. Specifically, we adopt the values $\sigma_{\pi N} = 20$, $30$ and $40$ MeV 
and use the respective number when tagging the NLO models. The choice $\sigma_{\pi N} = 30$ MeV 
was also used in our CS30 and NLO30 models applied to the $\bar{K}N$ system and discussed in 
\cite{2012CS}. To distinguish the model notation from our previous works we add a subscript $\eta$ 
for our present models fitted to the $\eta N$ related data. As we see from Table \ref{tab:fits} 
all three fits give a comparable level of data reproduction with the model NLO20$_{\eta}$ 
reaching only slightly better value of $\chi^{2}/dof$ that the NLO30$_{\eta}$ model. We also 
tried to perform the fits without restricting the energy region for the $I=3/2$ $\pi N$ amplitudes 
(to energies below $1450$ MeV) and without accounting effectively for the $\pi \pi N$ channel 
(a factor 1.2 in the $\eta N$ production cross section). With the parameters fixed to $\sigma_{\pi N} = 30$ 
we got $\chi^{2}/dof = 3.15$ then. This value drops significantly to 1.69 when we exclude 
the $I=3/2$ $\pi N$ data affected by the $\Delta(1620)$ resonance from the fit. A further improvement 
from 1.69 to the 1.46 reported for the NLO30$_{\eta}$ model is due to the effective accounting 
of the $\pi \pi N$ channel. Thus, we conclude that our treatment of the experimental data 
is reasonable and leads to their consistent reproduction.

\begin{table}[h]
\caption{The fit results and parameters of our NLO models. The inverse ranges $\alpha_{j}$ are in MeV, 
the NLO couplings $d$ in GeV$^{-1}$. The parameters of the KSW model \cite{1995KSWplb} are included  
for comparison.}
\begin{center}
\begin{tabular}{cc|cccc|cccc}
model          & $\chi^{2}/dof$ & $\alpha_{\pi N}$ & $\alpha_{\eta N}$ & $\alpha_{K\Lambda}$ & $\alpha_{K \Sigma}$ 
                          &      $d_D$       &        $d_F$      &       $d_0$         &     $d_1$  \\ \hline
NLO20$_{\eta}$ &     1.33       &        597       &       1293        &        256          &      1032       
                          &      2.062       &      -0.896       &      -2.279         &      3.528 \\
NLO30$_{\eta}$ &     1.46       &        538       &       1635        &        250          &       939       
                          &      1.981       &      -0.770       &      -2.452         &      3.940 \\
NLO40$_{\eta}$ &     1.95       &        508       &       2000        &        250          &       842       
                          &      1.863       &      -0.685       &      -2.608         &      4.366 \\
KSW            &   ---    &        573       &       776         &        776          &       776        
                          &      0.420       &      -0.410       &      -0.745         &     -0.380  
\end{tabular}
\end{center}
\label{tab:fits}
\end{table}

Concerning the KSW model we emphasize that the model is included in Table~\ref{tab:fits} only 
for a reference and comparison of the fitted parameters. The pertinent value of $\chi^{2}/dof$ 
is omitted in Table~\ref{tab:fits} since a direct comparison with the $\chi^{2}/dof$ values of 
our new fits is not meaningful. This is because not only the data set was different in \cite{1995KSWplb} 
from the one used here but a number of the fitted parameters $N_{par}$ and a number of observables $N_{obs}$ 
were different in Ref.~\cite{1995KSWplb} too. As anticipated we checked that the KSW model 
leads to unacceptably large $\chi^{2}/dof$ when calculated with the present fitting metodology 
applied to the whole energy interval starting from the $\pi N$ threshold. 

It is worth noting the differences in the fitted parameters. All our three models 
lead to a $\pi N$ inverse range $\alpha_{\pi N} \approx 550$ MeV in good agreement with 
the KSW value. Our models are characterized by very large inverse range in the $\eta N$ 
channel making the $\eta N$ system quite compact. The opposite can be said about the $K\Lambda$ 
channel while a moderate value of $\alpha_{K\Sigma} \approx 900$ MeV resulted for the $K\Sigma$ 
channel. We have no explanation for such vast differences between the involved meson-baryon 
systems. Though, in average our results are in line with a common value of the inverse range 
used for the three nonpionic channels in the KSW model. 

\begin{figure}[h]    
\centering
\includegraphics[width=\textwidth]{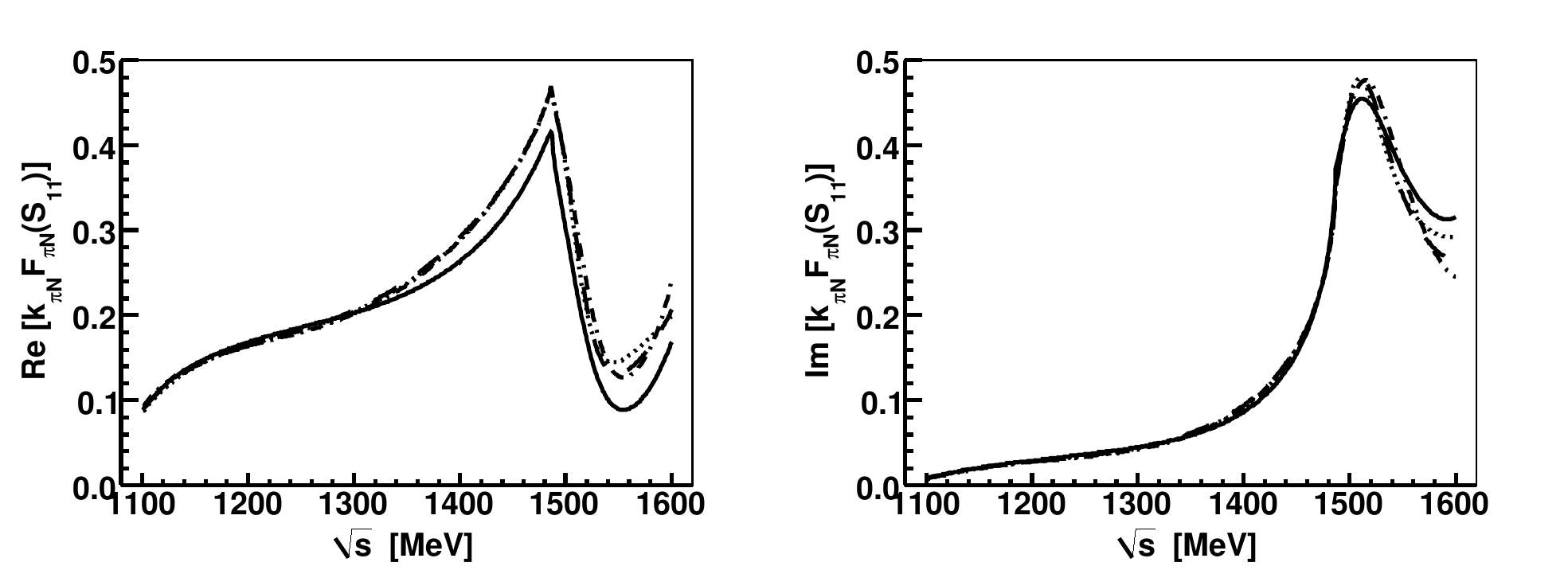}
\includegraphics[width=\textwidth]{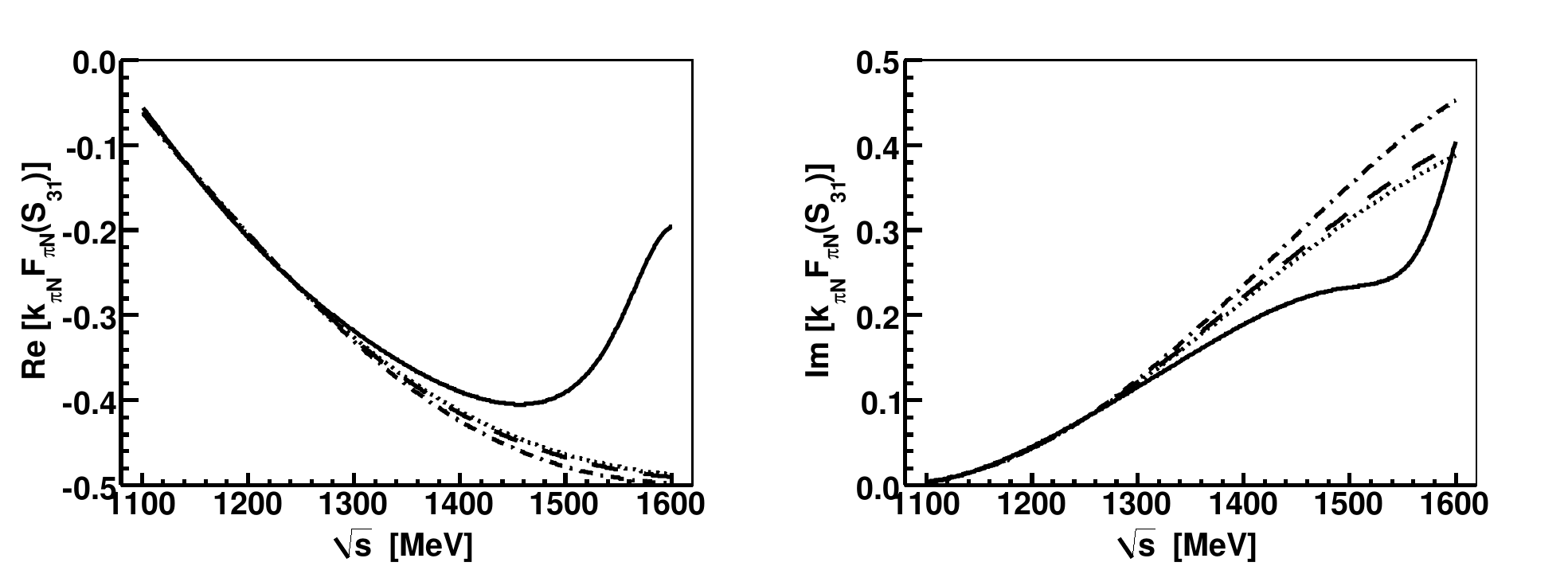}
\caption{The real (left panels) and imaginary (right panels) parts of the unitless 
$k_{\pi N}F_{\pi N}(S_{11})$ (top panels) and $k_{\pi N}F_{\pi N}(S_{31})$ (bottom panels) 
amplitudes generated by the NLO20$_{\eta}$ model (dotted line), NLO30$_{\eta}$ model (dashed line) 
and the NLO40$_{\eta}$ model (dot-dashed line). The continuous line represents 
the SAID partial wave solution.}
\label{fig:aPiN}
\end{figure}

The $d$-couplings obtained in our fits and shown in Table \ref{tab:fits} are quite large and completely 
at variance with the KSW model, that attempted to keep them close to those from $\bar{K}N$ studies 
\cite{1995KSWnpa} made earlier by the same authors, and with our own previous studies of the $\bar{K}N$ system
\cite{2010CS}, \cite{2012CS}. In principle, it seems natural to assume a common low energy constants 
for both $S=-1$ and $S=0$ strangeness sectors. However, as the $d$-couplings are the only Lagrangian 
couplings that are left free in our model, parts of them effectively make up for all contributions that 
are not accounted for at the NLO level. The higher order corrections omitted at the NLO level 
are different in the $S= 0$ and $S=-1$ sectors and apparently more important in the $S= 0$ sector 
than in the $S= -1$ one. In other words, the large values of our $d$-couplings indicate a worse 
convergence of the $\chi PT$ in the $S=0$ sector. This is also reflected by a commonly 
accepted observation that the chiral expansion works reasonably well for the $\bar{K}N$ system already 
at the LO order, at least in the energy region around the $\bar{K}N$ threshold. On the contrary, 
we have found that we could not achieve a satisfactory description of the $\pi N$ and $\eta N$ experimental 
data without introducing the NLO terms. We have also made an attempt to vary only the inverse ranges $\alpha_{j}$ 
while fixing the $d$-couplings to those established in our previous NLO30 fit to the $\bar{K}N$ data. 
The resulting $\chi^{2}/dof \approx 60$ is not acceptable and clearly indicates that 
the $S=-1$ and $S=0$ data are difficult to describe with a common set of low energy constants 
at the NLO level. The achievement of Ref.~\cite{1995KSWnpa} is a bit misleading in this sense 
as we found that the applicability of the KSW model is restricted only to a narrow interval of 
energies above the $\eta N$ threshold. We also note that quite different NLO couplings for the $S=0$ 
and $S= -1$ sectors were obtained in separate fits to the pertinent data by another group of authors 
in Refs.~\cite{2012MBM} and \cite{2013MM}, respectively. Of course, it would be best if a simultaneous 
fit to both the $\bar{K}N$ and $\eta N$ related data were attempted to find out whether a common set 
of LECs exists for both sectors (as unlikely this might seem), but this goes beyond the scope of our 
present work.

In Figure \ref{fig:aPiN} we demonstrate how well our models reproduce the $s$-wave $\pi N$ amplitudes 
from the SAID database. In the whole energy region all three models give equivalent $\pi N$ amplitudes 
with hardly noticeable differences among them, especially below the $\eta N$ threshold. The overall 
reproduction of the SAID amplitudes is quite good for the $S_{11}$ partial wave. Deviations 
from the SAID amplitudes observed for the $S_{31}$ amplitude at higher energies are caused 
by the $\Delta(1620)$ resonance (which is not accounted for in our model) and justify our omission 
of the affected $I=3/2$ data from the fits. We note that much larger variance between the calculated 
and SAID amplitudes is obtained when only the TW term is used in the effective 
meson-baryon potentials, see e.g.~Fig.~2 in Ref.~\cite{2002IOV}. On the other hand, the quality 
of the fits performed in Ref.~\cite{2012MBM} seem to be slightly better than ours, though the authors 
fitted a bit different set of experimental data and varied a larger number of model parameters. 
A qualitative comparison of predictions for the $\pi N$ amplitudes made in the current work 
and in Ref.~\cite{2012MBM} with those of Ref.~\cite{2002IOV} once again demonstrate 
the importanance of the NLO interaction terms, specifically at low energies between the $\pi N$ and 
$\eta N$ thresholds.

\begin{figure}[h]    
\centering
\includegraphics[width=0.75\textwidth]{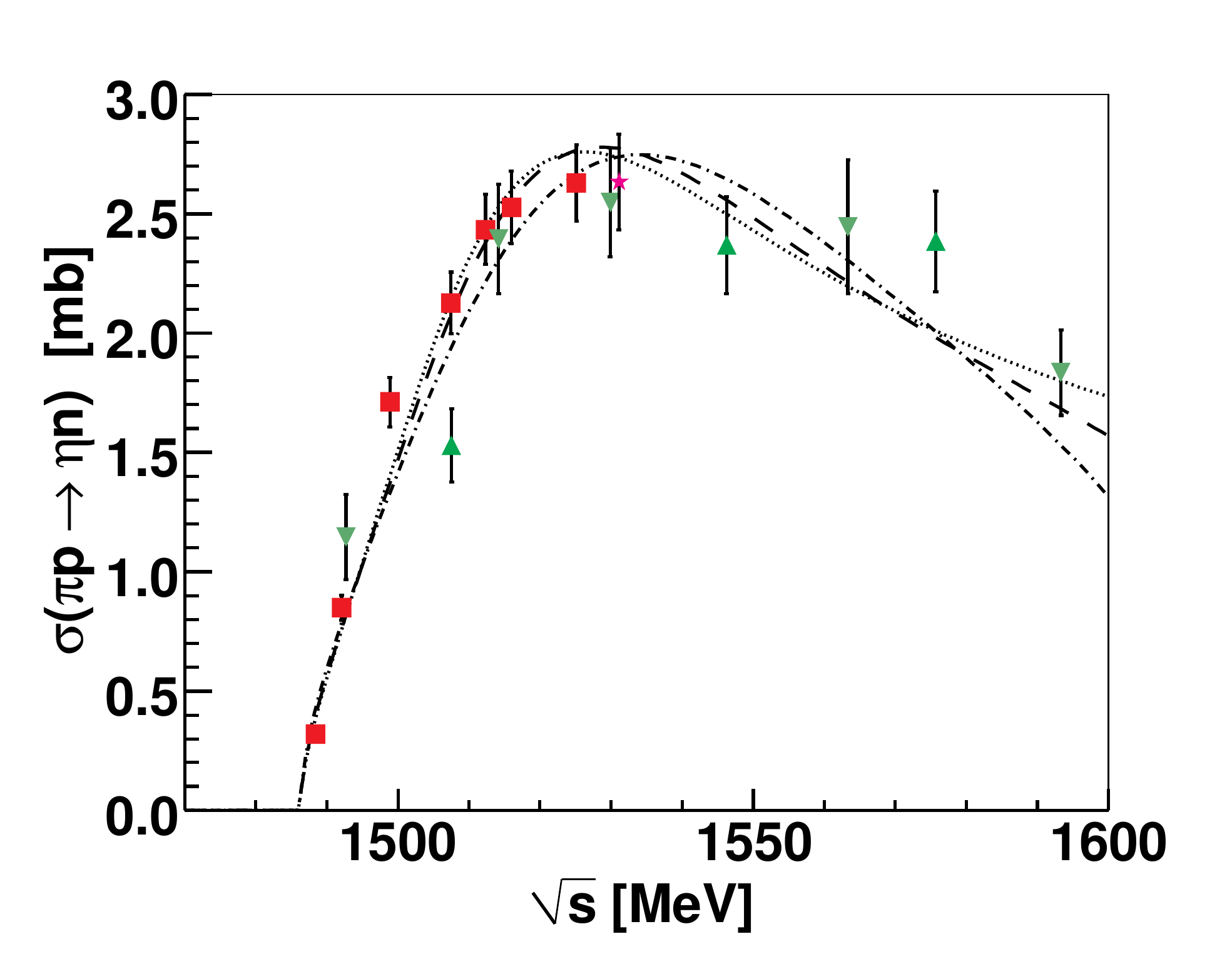}
\caption{A comparison of the model predictions for the $\pi^{-}p \rightarrow \eta n$ total cross section. 
The results obtained with the NLO20$_{\eta}$ model (dotted line), NLO30$_{\eta}$ model (dashed line) 
and the NLO40$_{\eta}$ model (dot-dashed line), are plotted together with the experimental data.}
\label{fig:etaNxs}
\end{figure}

In Figure \ref{fig:etaNxs} we present the calculated $\pi^{-}p \rightarrow \eta n$ total cross sections 
in comparison with the experimental data, the same ones as in Figure \ref{fig:KSW}. The agreement 
of the theoretical predictions with the experimental cross sections is quite good for all three models.
The models give very similar results for energies up to about 1580 MeV, then start to deviate. 
Apparently, a fit to higher energy data would be necessary to make distinction among them. 
However, this would require an introduction of higher partial waves and a proper treatment 
of the $\pi \pi N$ channel, tasks that go beyond restrictions of the present approach. 

The peak structure of the $\eta N$ production cross section relates to the $N^{\star}(1535)$ 
resonance and is nicely reproduced by our models. We have also noted that the peak position 
is closely related to the compactness of the $\eta N$ system resulting from our fits. 
When we fixed the inverse range $\alpha_{\eta N}$ at a smaller value (e.g. at 700 MeV) 
and fitted the remaining model parameters we were able to achieve a reasonably good overall 
agreement with the experimental data, though only on the expense of moving the $\eta N$ production 
maximum to higher energies, about 30 MeV off the one seen in Figure~\ref{fig:etaNxs}. We will 
come back to this feature when discussing the position of a pole related to the $N^{\star}(1535)$ 
resonance in Section \ref{sec:poles}.

\subsection{$\eta N$ amplitudes}
\label{sec:EtaN}

\begin{figure}[h]    
\centering
\includegraphics[width=\textwidth]{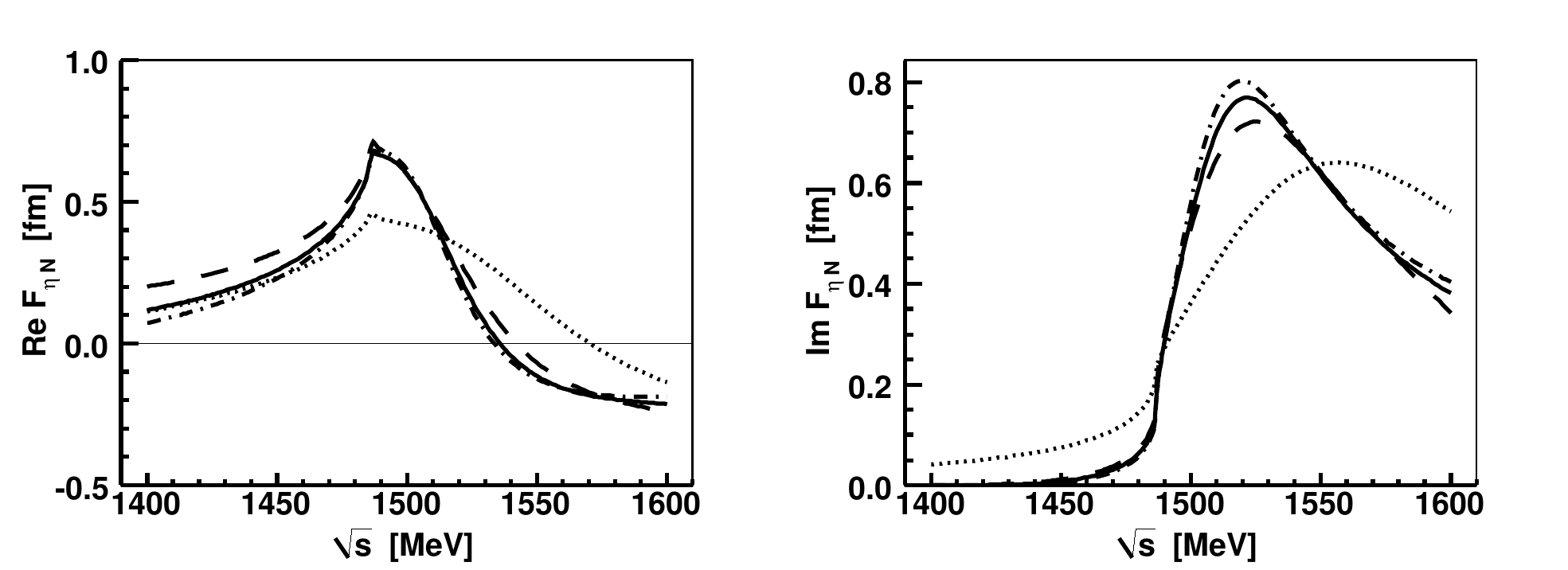}
\caption{The real (left panel) and imaginary (right panel) parts of the $\eta N$ amplitude generated 
by the KSW model (dotted line), the NLO20$_{\eta}$ model (dot-dashed line), NLO30$_{\eta}$ model 
(continuous line) and the NLO40$_{\eta}$ model (dashed line).}
\label{fig:aEtaN}
\end{figure}

The energy dependence of the $\eta N$ elastic amplitudes is visualized in Figure~\ref{fig:aEtaN}. 
All our three NLO models lead to very similar predictions for the $\eta N$ amplitudes, at some 
parts difficult to distinguish one from the other. This is in contrast to the predictions 
of the KSW model that lead to a smaller attraction at the $\eta N$ threshold and to much 
larger imaginary part of the amplitude below the threshold. The $\eta N$ scattering lenghts 
generated by the NLO30$_{\eta}$ and the KSW models are $a_{\eta N} = (0.67+{\rm i}\:0.20)$ fm and 
$(0.46+{\rm i}\:0.24)$ fm, respectively\footnote{The $\eta N$ scattering length reported 
in Ref.~\cite{1995KSWplb} is $a_{\eta N} = (0.68+{\rm i}\:0.24)$ fm having a real part at variance 
with our reproduction of the KSW results. Most likely, the value given in Ref.~\cite{1995KSWplb}
refers to their local potential, not to the KSW separable potential we reproduce and compare with here.}. 
Our $\Re a_{\eta N}$ values are also significantly larger than most of other predictions 
by chirally motivated models \cite{1995KSWplb}, \cite{1997KWW}, \cite{2002IOV}, 
\cite{2012MBM} and closer to the value of $\Re a_{\eta N} \approx 1$ fm established in $K$-matrix 
analyses of the $\pi N$ and $\eta N$ reaction data \cite{2005GW}, \cite{2005Arn}, \cite{2013SLM}. 
A similar value of $a_{\eta N} = (0.77+{\rm i}\:0.22)$ fm, quite close to our result, was 
obtained in Ref.~\cite{2001NR}.

As we already mentioned in Section \ref{sec:intro} the strength of the $\eta N$ threshold 
(and subthreshold) attraction was discussed in Ref.~\cite{2013FGM} in relation to a possible 
formation of the $\eta$-nuclear bound states. There, the authors found that the attraction  
provided by models that generate $\Re a_{\eta N} \lsim 0.5$ fm is not sufficient for forming 
the $\eta$-nuclear bound states in the lightest nuclei including $^{12}C$. Taking into 
consideration their results we estimate that our model attraction might be just sufficient 
to form the $1s_{\eta}$ nuclear bound states from $^{12}C$ on. However, a proper in-medium 
treatment of the $\eta N$ interaction requires a shift of the interaction energy 
to the subthreshold region as well as introduction of hadron selfenergies in the intermediate 
state propagator. The first effect was accounted for in Ref.~\cite{2013FGM} while 
an impact of the latter one is studied in another work prepared for publication \cite{2013CFGM}. 
The in-medium dressing of hadrons was also considered in an earlier work by Inoue and Oset \cite{2002IO} 
who reported an increased in-medium $\eta$-nuclear attraction at subthreshold energies. 
Although our own preliminary findings do not comply with their observations in this respect 
it is quite clear from both works, the \cite{2013CFGM} and \cite{2002IO} ones, 
that a proper treatment of the $\eta$-nuclear interaction depends on a reliable extrapolation 
of the $\eta N$ interaction to subthreshold energies. In this sense, it is incouraging 
to note a good agreement (at least for the real part of the $\eta N$ amplitude) at the subthreshold 
energies between our three models and the KSW one.

Our models predict a very sharp drop of the imaginary part of the $\eta N$ amplitude 
at the subthreshold energies. In Figure~\ref{fig:aEtaN} we see that the KSW model leads 
to significantly larger $\Im F_{\eta N}$ there. However, as the KSW model does not reproduce 
well the $\pi N$ amplitudes at energies below the $\eta N$ threshold we maintain 
that our results are more realistic within the model constrains. The small $\Im F_{\eta N}$ 
implies a small absorptivity of the $\eta$-nuclear optical potential. Though, one should also 
remember that models discussed here lack completely the three body $\pi \pi N$ channel. 
Its introduction should lead to a larger imaginary part of the elastic $\eta N$ amplitude 
and to a related increase of the imaginary part of the optical potential. In close resemblance 
to $\bar{K}$-nuclear studies \cite{2012FG}, \cite{2013FG} we also expect contributions 
to the optical potential, to both its real as well as imaginary parts, arising from meson 
interactions with nucleon pairs (and clusters in general).

\section{Dynamically generated resonances}
\label{sec:poles}

In Table \ref{tab:poles} we show the positions of the poles our models generate on two Riemann sheets 
that are connected with the physical region in the considered energy interval. The Riemann sheet (RS) connected 
to physical region by crossing the real axis between the $\eta N$ and $K \Lambda$ thresholds is denoted 
as [-,-,+,+] with the signs marking the signs of the imaginary parts of the meson-baryon CMS momenta in all four 
coupled channels (unphysical for the $\pi N$ and $\eta N$ channels and physical for the $K \Lambda$ 
and $K \Sigma$ ones). Similarly, the RS connected with physical region in between 
the $K\Lambda$ and $K\Sigma$ thresholds is denoted as [-,-,-,+]. We note that all three models 
give very similar predictions for the $z_1$ pole that can be assigned to the $N^{\star}(1535)$ 
resonance. The position of the $z_2$ pole is not so well determined by our models. For the NLO20$_{\eta}$ 
and NLO30$_{\eta}$ models we find the pole on the RS reached by crossing the real axis above the $K\Lambda$ 
threshold but the pole itself lies below it. Even then, it is still the pole that is closest to 
the physical region for energies above the $K\Lambda$ threshold. For this reason it is natural 
to assign it to the $N^{\star}(1650)$ resonance which was reported as a dynamically generated 
state in other chirally motivated coupled channels approaches to the $\eta N$ system \cite{2001NR}, 
\cite{2012MBM}. The difference between $\Re z_2$ and the experimentally established peak position 
of the $N^{\star}(1650)$ resonance can be explained by two factors. First of all the pole energy 
is shifted with respect to the peak energy due to an interference with background, similarly 
as it is so for the $N^{\star}(1535)$ pole $z_1$. Secondly, our models are based exclusively 
on the experimental data for energies below 1600 MeV. Thus, any model predictions for higher 
energies may be very unprecise quantitatively. It is also interesting to note that 
the pole positions of both poles reported in Table \ref{tab:poles} show a trend 
of increasing the pole energies (the real as well as imaginary parts) with the value 
of the $\sigma_{\pi N}$ term the model parameters are fixed to. Considering a possible shift 
of the pole to lower energies with respect to the cross section peak position, 
one can deduce that for $\sigma_{\pi N} \approx 35$ MeV the $z_2$ pole would be at about right 
place for an assignment to the $N^{\star}(1650)$ resonance.

\begin{table}[htb]
\caption{The positions of the $S_{11}$ poles generated by the models. The poles $z_1$ and $z_2$ 
are located on the [-,-,+,+] and [-,-,-,+] Rieman sheets, respectively. The last line shows 
PDG \cite{2012PDG} pole estimates for the $N^{\star}(1535)$ and 
$N^{\star}(1650)$ resonances.}

\begin{center}
\begin{tabular}{c|cc}
model              & $z_1$ [MeV] &  $z_2$ [MeV] \\[1mm] \hline 
NLO20$_{\eta}$     & 1502 - i33  &  1548 - i39  \\
NLO30$_{\eta}$     & 1503 - i37  &  1579 - i81  \\
NLO40$_{\eta}$     & 1504 - i48  &  1631 - i167 \\[1mm] \hline
PDG \cite{2012PDG} & 1510 - i85  &  1655 - i70  \\[1mm] \hline
\end{tabular}
\end{center}

\label{tab:poles}
\end{table}

When comparing our pole positions with the pole estimates by the Particle Data Group (PDG) \cite{2012PDG} 
one should keep in mind model ambiguities related to extending the resonance properties determined  
in various experiments to the complex energy plane. In Table \ref{tab:poles} we give only the average 
PDG estimates while the complete PDG listings provide pole positions determined by various authors 
that cover relatively broad region of energies around the PDG averages. The positions of our 
$z_1$ pole are in nice agreement with the PDG listings for the $N^{\star}(1535)$ with the imaginary 
part $\Im z_1$ provided by our NLO models at about the lower boundary of model predictions by other 
authors. Unfortunately, the same cannot be said about our predictions for the $N^{\star}(1650)$ 
pole, most likely due to reasons we gave above.

\begin{figure}[htb]    
\centering
\includegraphics[width=0.75\textwidth]{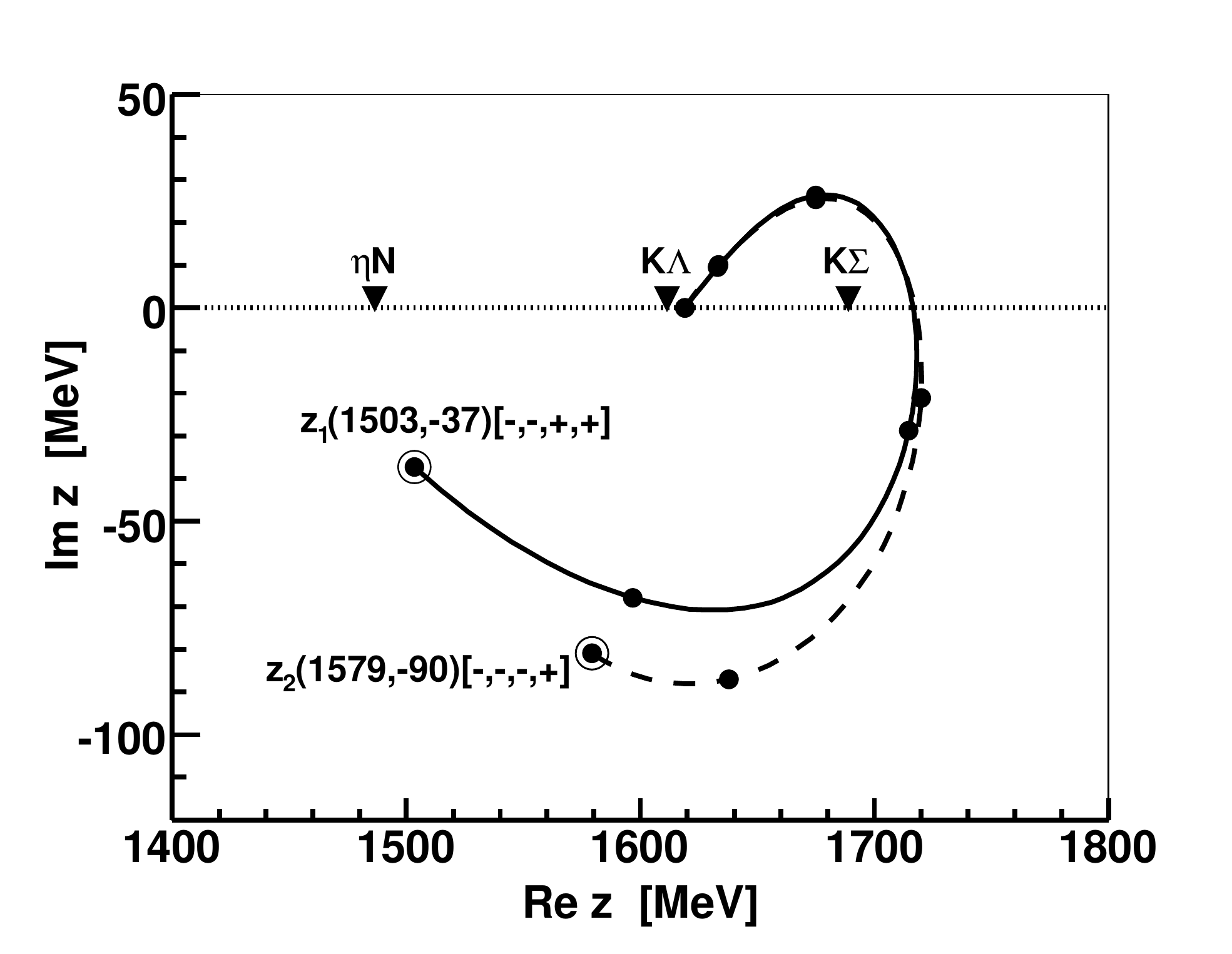}
\caption{Movement of the poles $z_1$ and $z_2$ upon gradually switching off the inter-channel couplings. 
The positions of the poles in a physical limit are encircled and marked by the labels that also denote 
the Riemann sheets the poles are located on. The small dots mark positions of the poles for the scaling 
factors from $x=0$ (zero coupling limit) to $x=1$ (physical limit) in steps of $0.2$.}
\label{fig:ZCL}
\end{figure}

The origin of resonances generated dynamically due to couplings between various meson-baryon channels 
can be traced to existence of poles in the zero coupling limit (ZCL), a hypothetical situation in which 
all inter-channel couplings are set to zero and the poles may persist only in those single channels that 
have nonzero diagonal couplings $C_{jj}$ \cite{1964ET}, \cite{1989PG}. This is also a case 
for the $\Lambda(1405)$ resonance which appears to be composed of two very close resonances, 
one of them related to a bound state in the $\bar{K}N$ channel, the other to a resonance 
in the $\pi \Sigma$ channel. There, the $\bar{K}N$-$\pi \Sigma$ inter-channel dynamics 
move the poles from their positions in the ZCL to those predicted by the models 
in the physical limit \cite{2008HW}, \cite{2013CS}. The situation is similar in 
the strangeness $S=0$ sector for the $I=1/2$ isospin where only the $\pi N$ and $K\Sigma$ channels 
have nonzero diagonal couplings. Since the $\pi N$ threshold is too far below the $\eta N$ one 
we anticipate that the dynamically generated poles $z_1$ and $z_2$ emerge from the $K\Sigma$ 
system. Indeed, this is a case as we demonstrate in Figure \ref{fig:ZCL} made with the NLO30$_{\eta}$ 
model. The diagonal $K\Sigma$ coupling is strong enough to generate a virtual state at an energy 
about 70 MeV below threshold. In the ZCL this gives us a pole position at 
the real axis on a RS that is unphysical in the $K\Sigma$ channel. As soon as the inter-channel 
couplings switch on this pole departs from the real axis and can start moving on any of 
the Riemann sheets that keep the minus sign for the $K\Sigma$ channel. In Figure \ref{fig:ZCL} 
we follow the movements of the poles on the Rieman sheets [+,+,-,-] (continuous line, $z_1$ pole) 
and [+,+,+,-] (dashed line, $z_2$ pole) in the upper half of the complex energy plane. 
The pole trajectories show the pole positions as we gradually increase a scaling factor $x$ 
that is applied to the non-diagonal inter-channel couplings $C_{ij}$ from $x=0$ 
(zero couling limit) to $x=1$ (physical limit). The dots mark the positions of the poles 
for $x=0$, $x=0.2$, ..., $x=1$ with the last point showing the final pole positions 
at full physical couplings, those given in Table \ref{tab:poles} for the NLO30$_{\eta}$ model.
The figure demonstrates that both poles, $z_1$ and $z_2$, evolve from the same origin 
and that they are mutually shadow poles. For small values of the scaling factor $x \lsim 0.5$ 
the pole positions on both Riemann sheets remain very close to each other and the two trajectories 
are hard to separate. The trajectories cross the real axis at energies above the highest 
meson-baryon threshold, so the poles continue their movements in the lower 
part of complex energy plane on Riemann sheets with all the signs reversed, the $z_1$ pole 
on the [-,-,+,+] RS and the $z_2$ pole on the [-,-,-,+] RS. In this part of the figure 
the trajectories are already clearly separated and the final pole positions for $x=1$ 
(in the physical limit) are quite different.

To complete our analysis of pole movements from the ZCL we add that there are also shadow poles which 
emerge on the [+,-,+,-] and [+,-,-,-] Rieman sheets for small values of $x$. Their trajectories cross 
the real axis twice, first about 10 MeV below the $K\Sigma$ threshold, then again immediately above 
the threshold, so the poles evolve on the [+,-,+,+] and [+,-,-,+] Rieman sheets for $x \gsim 0.6$, 
respectively. The final positions of those poles for $x=1$ are at the energies $z_3 = (1528 - {\rm i}19)$ MeV 
on the [+,-,+,+] RS and at $z_4 = (1634 - {\rm i}55)$ MeV on the [+,-,-,+] RS. Though the energies are 
in agreement with $N^{\star}(1535)$ and $N^{\star}(1650)$ poles expectations, the [+,-,-,+] RS is too far 
from physical region for the $z_4$ pole to have any impact on physical observables. The $z_3$ pole on 
the [+,-,+,+] RS is relatively close to the $\eta N$ threshold and not that far from physical region, 
but it plays only a secondary role to the pole $z_1$ on the [-,-,+,+] RS that is connected with 
the physical RS in between the $\eta N$ and $K\Lambda$ thresholds. Finally, we note in passing
that no poles emerge from the ZCL on the other Rieman sheets (e.g.~on the [-,-,-,-] RS) that are unphysical 
in the $K\Sigma$ channel. Thus, the two pole trajectories depicted in Figure \ref{fig:ZCL} are 
the only two that bring the poles to positions which are physically relevant.

We would like to point out that the model does not leave any room for more dynamically generated 
resonances with $I=1/2$ in the energy region above the $\eta N$ threshold. If there is a resonance on any 
RS connected to the physical region in that area it must originate from the virtual state in the $K\Sigma$ 
channel found in the ZCL. We checked that the other channel with nonzero diagonal coupling, 
the $\pi N$ one, provides a resonance too far from the real axis which is not moved any closer due to 
inter-channel dynamics. Thus, provided that the resonances $N^{\star}(1535)$ and $N^{\star}(1650)$ are both
generated dynamically, they should be viewed as two manifestations of only one substance that emerge 
as reflections of two shadow poles on two different Rieman sheets. The observable physical quantities 
(cross sections) are always affected by the pole that is closer to a given energy in the physical 
region (at the real axis), or that couples more strongly to a given physical state. It is also 
known that sometimes two shadow poles may have a comparable bearing on a physical observable, 
so both of them will affect it.

The pole positions can be found as solutions (for the complex energy variable $z$) 
of an equation that sets to zero the determinant of the inverse of the $F$ matrix, 
${\rm det}|F^{-1}(z)| = 0$ \cite{2013CS}. In the ZCL the condition for a pole 
in the $n$-th channel is equivalent to solving an equation $1/v_{nn}(z) = G_{n}(z)$ 
which leads to
\beq
\frac{4\pi f_{n}^{2}}{C_{nn}(z)}\: \frac{z}{M_n}
+ \frac{(\alpha_n + {\rm i}k_n)^{2}}{2\alpha_{n}}\:[g_{n}(k_{n})]^{2} = 0                   
\;\;\; .
\eeq{eq:ZCL}
This equation clearly shows that the position of the pole in the ZCL depends on the strength 
of the coupling $C_{nn}$ and on the value of the inverse range $\alpha_n$ that enters the expression 
for the form factor $g_{n}(k_{n})$ as well. This was already realized in Ref.~\cite{1995KSWplb} 
where a formation of a resonance state was linked to the magnitude of the range parameter, 
see Table 1 there. Our Eq.~(\ref{eq:ZCL}) provides a qualitative understanding of the relation. 
As the pole is moved from the ZCL due to inter-channel couplings the inverse ranges of the other 
channels get involved too and the situation becomes much more complicated.  
The final position of the pole in the physical limit is affected by many factors including 
the energy dependent inter-channel couplings $C_{ij}(z)$, thus it is difficult to establish 
a clear link between the inverse ranges and a position of the pole. For the NLO30$_{\eta}$ model 
we checked numerically that the $z_1$ pole related to $N^{\star}(1535)$ moves to higher energies 
and further from the real axis with a decreasing value of $\alpha_{\eta N}$. 

With the pole positions established one can also determine the couplings of involved channels 
to the pertinent resonant states. Here we follow Ref.~\cite{1972Tay} and express the transition 
amplitude in a vinicity of the complex pole energy $z_{R} = E_R - {\rm i}\Gamma_R /2$ as
\beq
F_{ij}(z) = F^{\rm BG}_{ij}(z) -\frac{1}{2(k_{i}k_{j})^{1/2}} \frac{b_{i} b_{j}}{z-z_{R}} \;\;\; ,
\eeq{eq:Fres2}
where the non-resonant background contribution $F^{\rm BG}$ and the dependence of the resonant 
part on the on-shell momenta $k_j$ are shown explicitly. The symmetric matrix $b_{i} b_{j}$ 
has rank 1 and the complex couplings $b_j$ can be determined from the residui of elastic 
scattering amplitudes calculated at the pole energy. They are related to the partial widths 
\beq
\Gamma_j = \mid b_j \mid ^{2} = \lim_{z \to z_R} \mid 2k_{j}(z-z_R)F_{jj}(z) \mid
\eeq{eq:G_i}
that refer to the probability of the resonant state decaying into the $j$-th channel. 
In reality, a resonant state can decay only into channels that are open at a given 
energy, so our Eq.~(\ref{eq:G_i}) does not make a good sense for channels with 
thresholds much higher than the pole energy $E_R$. However, in a case of a broad 
resonance (when the pole is not close to the real axis) the branching ratios calculated 
at the complex pole energy $z_R$ may differ significantly from the branching ratios 
established in experiment and assigned to the resonance peak energy. In addition, 
a tail of a resonance can also contribute to experimentaly measured branching ratio 
even in a situation when the pole energy $E_R$ (or a resonance peak) is found below 
the channel threshold. Finally, from the experimental point of view, it is very difficult 
(if not impossible) to establish the partial widths independently of a formation process 
that is particular for a given measurement. Thus, any theoretical partial widths calculated 
at the pole position are only approximately related to those established experimentaly. 
Keeping all that in our mind we have applied Eq.~(\ref{eq:G_i}) not only to channels 
open at the pole energy $E_R$ but to the closed ones as well.

In Table \ref{tab:BR} we present the unitless branching ratios defined as
\beq
BR_{j} = \frac{\Gamma_j}{\Gamma_R} = \frac{\Gamma_j}{2\; \Im z_R}
\eeq{eq:delta}
When looking at the numbers in the table one notices that the standardly anticipated unitarity 
relation $\sum_j BR_j = 1$ is sizebly violated, even when the summation index runs only over 
the open channels. In view of the ambiguities discussed above and because the peak 
of the $N^{\star}(1650)$ resonance lies above the $K\Lambda$ threshold we consider 
the $K\Lambda$ channel as open for the $z_2$ pole, though the pole is located below  
the threshold for the NLO20$_{\eta}$ and NLO30$_{\eta}$ models. For both the $z_1$ 
and $z_2$ poles the sums of the open channels branching ratios differ from unity 
by as much as $20\%$. However, in fact the relation should hold only for an isolated pole 
of the Breit-Wigner type and in absence of any background that would interfere with 
the resonance. There is no way to say whether the dynamically generated poles can 
be reliably approximated by the Breit-Wigner formula. Clearly, there is a background 
due to a distant presence of the pole related to the $\pi N$ channel 
and the branch cuts at the real axis that open at the channel threshold have some 
effect too. Thus, it is not surprising that the discussed relation holds just approximately 
even if we sum only over the open channels. The large values of the partial widths 
reported for the closed channels do not have a physical meaning and they reflect 
large magnitudes of the pertinent complex CMS momenta $k_j$. In fact, the effective   
couplings $\mid b_j \mid^{2} / (2\mid k_j \mid ) = \Gamma_{j} / (2\mid k_j \mid )$
are of a natural size.

\begin{table}
\caption{Calculated branching ratios for the poles $z_1$ and $z_2$ related to the $N^{\star}(1535)$ 
and $N^{\star}(1650)$ resonances, respectively. The last line shows the branching ratios 
estimated by the PDG \cite{2012PDG}.}
\label{tab:BR}     
\begin{center}
\begin{tabular}{c|cccc|cccc}
\hline\noalign{\smallskip}
         & \multicolumn{4}{c}{$z_1$ pole}              & \multicolumn{4}{c}{$z_2$ pole} \\
  model  & $\pi N$ & $\eta N$ & $K\Lambda$ & $K\Sigma$ & $\pi N$ & $\eta N$ & $K\Lambda$ & $K\Sigma$  \\
\noalign{\smallskip}\hline\noalign{\smallskip}
NLO20$_{\eta}$     & 0.33 & 0.70 & 7.92 & 11.4 & 0.40 & 0.26 & 0.14 & 8.27 \\
NLO30$_{\eta}$     & 0.36 & 0.73 & 19.9 & 16.7 & 0.35 & 0.15 & 0.15 & 4.46 \\
NLO40$_{\eta}$     & 0.41 & 0.82 & 35.5 & 26.5 & 0.81 & 0.19 & 0.21 & 6.75 \\
\noalign{\smallskip}\hline
PDG \cite{2012PDG} & 0.45 & 0.42 &  --- &  --- & 0.70 & 0.10 & 0.03 &  --- \\
\noalign{\smallskip}\hline
\end{tabular}
\end{center}
\end{table}

Similarly as for the pole positions we show only the PDG average estimates in Table \ref{tab:BR} 
rather then the whole intervals reported by various authors and listed by the PDG \cite{2012PDG}. 
As we already argued the branching ratios calculated from the amplitude residui at the pole positions 
may be quite different from the experimentaly determined ones. Thus, the comparison of our model 
predictions with the PDG estimates is presented here merely to illustrate how large the differences 
may be. It is still interesting to note the quite large BR$_{\eta N}$ predicted by our NLO models 
for the $N^{\star}(1535)$ resonance. Besides the reasons stated above another factor influencing 
this may be related to an ommission of the $\pi \pi N$ decays in our model that are effectively 
accounted for by $N^{\star}(1535)$ decays into the $\eta N$ channel.

\section{Summary}
\label{sec:sum}

In summary, we have presented new fits of the $\pi N$ and $\eta N$ data for the partial $s$-wave 
including CMS energies from the $\pi N$ threshold up to $1600$ MeV. The resulting chirally motivated 
meson-baryon potentials lead to the $\eta N$ scattering length $a_{\eta N} \approx (0.7+{\rm i}\:0.2)$ fm 
corresponding to much stronger attraction than generated by other chiral models. 
The separable character of our model with its natural off-shell extrapolation makes it suitable 
for in-medium applications, specifically for studies of possibly bound $\eta$-nuclear states. 
The energy dependence of our $\eta N$ amplitudes at subthreshold energies required in such studies 
is consistent with earlier predictions by other authors, though the imaginary part of our $\eta N$ 
amplitude may fall too fast when going subthreshold. If the same feature was preserved for the 
in-medium amplitude, the predicted $\eta$-nuclear $1s_{\Lambda}$ bound states would have 
unrealistically small absorption widths. This effect arises due to a restriction of our model 
to two body interactions, particularly due to an omission of the $\pi \pi N$ channel, the only 
inelastic channel in $\pi N$ induced reactions that opens below the $\eta N$ threshold. 
We leave an implementation of the $\pi \pi N$ channel for a future.

In the discussed region of energies the $\pi N$-$\eta N$ interactions are strongly affected 
by the $N^{\star}(1535)$ and $N^{\star}(1650)$ resonances, both of them generated dynamically 
by our model. We have found that both states have the same origin that can be traced 
to the $K \Sigma$ virtual state, established on the real axis below the $K \Sigma$ threshold 
on the unphysical RS when the inter-channel couplings are switched off. 

In general, the present analysis of $\eta N$ interactions compliments our earlier studies 
of the $\bar{K}N$ system which we did in the same coupled channels framework that relies 
on chiral dynamics. In both situations we have vitnessed a strong energy dependence 
of the elastic meson-baryon amplitudes at energies close to the threshold and particularly 
below the threshold. While the dynamics of the $\bar{K}N$ system is driven by the $\Lambda(1405)$ 
resonance the low energy $\eta N$ interactions are strongly affected by the $N^{\star}(1535)$ 
resonance. The discussions of both systems with the same methodology provide us with a better 
control of the related models and give us new insights on general aspects of meson-baryon 
interactions in the nonperturbative regime.

\section*{Acknowledgement}

A.~C. would like to thank E.~Friedman and A.~Gal for their warm hospitality and stimulating discussions 
during his stay at the Hebrew University in Jerusalem where part of the work was completed. 
The authors also acknowledge a help by R.~Workman who provided us with the source data on the $\eta n$ 
production icluded in the SAID database. 
The work of A.~C. was supported by the Grant Agency of Czech Republic, Grant No.~P203/12/2126.

\section*{Appendix}

The central piece of our pseudopotentials Eq.~(\ref{eq:vij}) is the coupling matrix $C_{ij}$ 
which form is derived from the SU(3) chiral symmetry and reflects the structure of the underlying 
chiral Lagrangian. Considering the leading and next-to-leading orders of the chiral expansion 
and making a projection to the $s$-wave the couplings can be expressed in a form \cite{1995KSWnpa}, 
\cite{2010CS}, \cite{2012CS}
\beqa
\!\!\!\! C_{ij}(\sqrt{s}) \!  
                & \! = \! & \! - C_{i j}^{\rm (TW)} (2\sqrt{s} \!-\! M_{i} \!-\! M_{j})/4 + 
                               C_{i j}^{(u)} \frac{1}{M_{0}} 
                               \left( - k_{i}^{2} -\! k_{j}^{2} +\! 
                                       \frac{1}{3}\frac{k_{i}^{2}\, k_{j}^{2}}{m_{i}\, m_{j}} 
                               \right) + \CR[.5em]
                &         & \! + C_{i j}^{(EE)} \, E_{i} E_{j} 
                               + C_{i j}^{(mm)} \:\!\left( m_{i}^{2} + m_{j}^{2} \right) +
                                         \CR[.5em] 
                &         & \! + C_{i j}^{(\chi {\rm b})} \:\!\left( m_{K}^{2} - m_{\pi}^{2} \right)
                               + C_{i j}^{({\rm GOv})} \! 
                                    \left(
                                    m_{\eta}^{2} -\frac{4}{3} \;\! m_{K}^{2} +\frac{1}{3} \;\! m_{\pi}^{2}
                                    \right) \; ,
\eeqa{eq:CijA}
where $m_{j}$, $E_j$ and $k_j$ denote the meson mass, CMS energy and momenta in channel $j$, respectively, 
and $M_0$ stands for the baryon mass in the chiral limit. The general structure of the coefficients 
$C_{ij}$ follows closely the one we adopted in our earlier studies of the $\bar{K}N$ interactions 
\cite{2010CS}, \cite{2012CS}. The terms marked by the superscripts "WT" and "u" 
correspond to the leading TW contact interaction and to the crossed Born amplitude 
represented by the diagrams a) and c) in Figure \ref{fig:FD}, respectively. 
The remaining parts contribute to the contact interaction in the next-to-leading order 
visualized by diagram d) in the figure. There, the last two terms on the third line 
in Eq.~(\ref{eq:CijA}) represent an explicit breaking of the chiral symmetry (the ``$\chi {\rm b}$" term) 
and a violation of the Gell-Mann--Okubo formula (the ``GOv" term).

\begin{table}[htb]
\label{tab:CTW}
\centering
\caption{The $C_{i j}^{\rm (TW)}$ coefficients for the $S=0$, $I=1/2$ and $I=3/2$ channels, 
$C_{j i}^{\rm (TW)} \!= C_{i j}^{\rm (TW)}$.}
\begin{tabular}{c|cccc|cc}
    \MySepRule
      & $ \pi N$                                         
      & $ \eta N$                                        
      & $ K \Lambda$                                     
      & $ K \Sigma$                                      
      & $ \pi N$                                         
      & $ K \Sigma$                           \\[\MySep] 
    \hline
     $\pi N$                          \rule{0em}{2.5\MySep}  
      & 2
      & 0
      & $\frac{3}{2}$
      & $-\frac{1}{2}$                            
      & 0
      & 0                                         \\[\MySep]
    $\eta N$ 
      &  
      & 0
      & $-\frac{3}{2}$
      & $-\frac{3}{2}$                           
      & 0
      & 0                                         \\[\MySep]  
    $K \Lambda$ 
      &
      & 
      & 0
      & 0                                         
      & 0
      & 0                                         \\[\MySep]
    $K \Sigma$  
      & 
      & 
      & 
      & 2
      & 0
      & 0                                         \\[\MySep]
    \hline 
     $\pi N$                            \rule{0em}{2.5\MySep}
      &
      & 
      &  
      &                                           
      & -1 
      & -1                                        \\[\MySep]
    $K \Sigma$  
      & 
      & 
      & 
      & 
      & 
      & -1
\end{tabular}
\end{table}

The reader may wonder why a direct Born term 
represented by diagram b) in Figure \ref{fig:FD} does not contribute to Eq.~(\ref{eq:CijA}). 
The reason lies in the way the direct and crossed Born diagrams are treated when the chiral 
Lagrangian is formed. Unlike in \cite{2010CS} which was based on the approach adopted in \cite{1995KSWnpa} 
here we follow the prescription used in \cite{2012CS} and based on Refs.~\cite{2000Fet}, \cite{2000FMMS}. 
Although both ways lead to equivalent form (up to an ${\cal O}(q^{2})$ order) of meson-baryon 
chiral amplitudes with only some rearrangement of the NLO coupling parameters, the latter approach 
is prefered for technical reasons. The advantage of this approach is that the $s$-wave projection 
of the direct Born term is exactly zero, so the pertinent contribution can be skipped 
in Eq.~(\ref{eq:CijA}). Both the direct and crossed Born terms are omitted whenever 
the KSW model of Ref.~\cite{1995KSWplb} is used in our calculations, in agreement with 
the approach adopted there.

\begin{table}[htb]
\label{tab:Cu}
  \centering
  \caption{The $C_{i j}^{(u)}$ coefficients for the $S=0$, $I=1/2$ and $I=3/2$ channels, 
      $C_{j i}^{(u)} \!= C_{i j}^{(u)}$. To shorten the length of the coefficients we denote 
      ${\cal U}[a,b] = \DxD + a\DxF + b\FxF$.
              }
  \vspace{0.75cm}

\begin{tabular}{c|cccc|cc}
    & $ \pi N$                                           
    & $ \eta \:\! n$                                     
    & $ K \:\!\! \it\Lambda$                             
    & $ K \:\!\! {\it\Sigma}$                            
    & $ \pi N$                                           
    & $ K \:\!\! {\it\Sigma}$                 \\[\MySep] 
  \hline
    \rule{0em}{2.75\MySep}
          $\pi N$ 
          \rule{0em}{2.75\MySep}                        
      & $- \frac{1}{4} {\cal U}[2,1]$
      & $\frac{1}{4} {\cal U}[-2,-3]$
      & $- \frac{1}{2} {\cal U}[-1,0]$
      & $\frac{1}{6} {\cal U}[-3,6]$
      & 0
      & 0                                         \\[\MySep]
    $ \eta \:\! n$
      &  
      & $\frac{1}{12} {\cal U}[-6,9]$
      & $\frac{1}{6} {\cal U}[3,0]$
      & $\frac{1}{2} {\cal U}[-1,0]$
      & 0
      & 0                                         \\[\MySep]
    $ K \:\!\! \it\Lambda$
      &
      & 
      & $\frac{1}{12} {\cal U}[-6,9]$
      & $- \frac{1}{4} {\cal U}[-2,-3]$
      & 0
      & 0                                         \\[\MySep]
    $ K \:\!\! {\it\Sigma}$ 
      &
      &
      &
      & $- \frac{1}{4} {\cal U}[2,1]$
      & 0
      & 0                                         \\[\MySep]
  \hline
    \rule{0em}{2.75\MySep}
          $ \pi N$ 
          \rule{0em}{2.75\MySep}
      & 
      &
      &
      &
      & $\frac{1}{2} {\cal U}[2,1]$
      & $- \frac{1}{6} {\cal U}[6,-3]$   \\[\MySep]
    $ K \:\!\! {\it\Sigma}$ 
      &
      &
      &
      &
      &
      & $\frac{1}{2} {\cal U}[2,1]$
  \end{tabular}
\end{table}

The coefficients $C_{i j}^{\rm (.)}$ of Eq.~(\ref{eq:CijA}) are combinations of low energy constants, 
the couplings that determine strengths of pertinent contributions to the chiral Lagrangian. 
The coefficients are presented in Tables 4 -- 10, each 
of them splitted in two sectors that do not couple one to the other. The first sector is composed 
of four channels ($\pi N$, $\eta N$, $K\Lambda$ and $K\Sigma$) related to the isospin $I=1/2$, the second 
sector is represented by two $I=3/2$ channels ($\pi N$ and $K\Sigma$). Since the coupling matrices are symmetric, 
$C_{j i}^{\rm (.)} = \, C_{i j}^{\rm (.)}$, we show only the terms above the diagonal. We also split 
the $ C_{i j}^{(EE)}$ coefficients in two parts, $ C_{i j}^{(EE)} = C_{i j}^{(EE1)} + C_{i j}^{(EE2)}$, 
as a single table would be too large to fit a page. We have checked that our $C_{ij}^{\rm (.)}$  
matrices reproduce exactly the couplings given in Apendices of Refs.~\cite{1995KSWplb} and \cite{1997KWW} 
with the exception of an opposite overall sign applied to coefficients $C_{12}$, $C_{13}$ and $C_{14}$. 
This discrepancy is related to different phase conventions when splitting the physical $\pi N$ states 
into $I=1/2$ and $I=3/2$ components and it has no impact whatsever on the calculated observables.

\begin{table}[htb]
\label{tab:EE1}
  \centering
  \caption{The $C_{i j}^{(EE1)}$ coefficients for the $S=0$, $I=1/2$ and $I=3/2$ channels, 
      $C_{j i}^{(EE1)} \!= C_{i j}^{(EE1)}$.}
  \vspace{0.75cm}

\begin{tabular}{c|cccc|cc}
    & $\pi N$                                            
    & $\eta N$                                           
    & $K\Lambda$                                         
    & $K\Sigma$                                          
    & $\pi N$                                            
    & $K\Sigma$                               \\[\MySep] 
  \hline
    \rule{0em}{2.75\MySep}
          $\pi N$
    \rule{0em}{2.75\MySep}                        
      & $-\DeDe -\DeFe$ 
      & $\DeDe +\DeFe$
      & $-\frac{1}{2}\DeDe -\frac{3}{2}\DeFe$
      & $-\frac{1}{2}\DeDe +\frac{1}{2}\DeFe$
      & 0
      & 0                                         \\[\MySep]
    $\eta N$
      &  
      & $-\frac{5}{3}\DeDe +\DeFe$
      & $-\frac{1}{6}\DeDe -\frac{1}{2}\DeFe$
      & $\frac{1}{2}\DeDe -\frac{1}{2}\DeFe$
      & 0
      & 0                                         \\[\MySep]
    $K\Lambda$
      &
      & 
      & $-\frac{5}{3}\DeDe$
      & $-\DeDe$
      & 0
      & 0                                         \\[\MySep]
    $K\Sigma$
      &
      &
      &
      & $-\DeDe +2\DeFe$
      & 0
      & 0                                         \\[\MySep]
  \hline
    \rule{0em}{2.75\MySep}
          $\pi N$ 
          \rule{0em}{2.75\MySep}
      & 
      &
      &
      &
      & $-\DeDe -\DeFe$
      & $-\DeDe +\DeFe$                           \\[\MySep]
    $K\Sigma$
      &
      &
      &
      &
      &
      & $-\DeDe -\DeFe$
  \end{tabular}
\end{table}

\begin{table}[htb]
\label{tab:CEE2}
  \centering
  \caption{The $C_{i j}^{(EE2)}$ coefficients for the $S=0$, $I=1/2$ and $I=3/2$ channels, 
      $C_{j i}^{(EE2)} \!= C_{i j}^{(EE2)}$.}
  \vspace{0.75cm}

\begin{tabular}{c|cccc|cc}
    & $\pi N$                                            
    & $\eta N$                                           
    & $K\Lambda$                                         
    & $K\Sigma$                                          
    & $\pi N$                                            
    & $K\Sigma$                               \\[\MySep] 
  \hline
    \rule{0em}{2.75\MySep}
          $\pi N$
    \rule{0em}{2.75\MySep}                        
      & $-2\DeNul$ %
      & $-\DeII$
      & $\frac{1}{2} \DeII$
      & $\DeI + \frac{3}{2} \DeII$
      & 0
      & 0                                         \\[\MySep]
    $\eta N$
      &  
      & $- 2\DeNul +\frac{2}{3} \DeII$
      & $- \DeI + \frac{5}{6} \DeII$
      & $- \frac{1}{2} \DeII$
      & 0
      & 0                                         \\[\MySep]
    $K\Lambda$
      &
      & 
      & $-2\DeNul +\frac{2}{3}\DeII$
      & $\DeII$
      & 0
      & 0                                         \\[\MySep]
    $K\Sigma$
      &
      &
      &
      & $-2\DeNul$
      & 0
      & 0                                         \\[\MySep]
  \hline
    \rule{0em}{2.75\MySep}
          $\pi N$ 
          \rule{0em}{2.75\MySep}
      & 
      &
      &
      &
      & $-2\DeNul$
      & $- \DeI$                                  \\[\MySep]
    $K\Sigma$
      &
      &
      &
      &
      &
      & $-2\DeNul$
  \end{tabular}
\end{table}

\begin{table}
  \centering
  \caption{The $C_{i j}^{(mm)}$ coefficients for the $S=0$, $I=1/2$ and $I=3/2$ channels, 
      $C_{j i}^{(mm)} = C_{i j}^{(mm)}$.}
  \vspace{0.75cm}

\begin{tabular}{c|cccc|cc}
    & $\pi N$                                            
    & $\eta N$                                           
    & $K\Lambda$                                         
    & $K\Sigma$                                          
    & $\pi N$                                            
    & $K\Sigma$                               \\[\MySep] 
  \hline
    \rule{0em}{2.75\MySep}
          $\pi N$
          \rule{0em}{2.75\MySep}
      & $2 \BeNul \!+\! \BeDe \!+\! \BeFe$ 
      & $- \BeDe \!-\! \BeFe$
      & $\frac{1}{2}\BeDe \!+\! \frac{3}{2}\BeFe$
      & $\frac{1}{2}\BeDe \!-\! \frac{1}{2}\BeFe$
      & 0
      & 0                                         \\[\MySep]
    $ \eta N$
      &  
      & $2 \BeNul \!+\! \frac{5}{3}\BeDe \!-\! \BeFe$
      & $\frac{1}{6}\BeDe \!+\! \frac{1}{2}\BeFe$
      & $- \frac{1}{2}\BeDe \!+\! \frac{1}{2}\BeFe$
      & 0
      & 0                                         \\[\MySep]
    $ K \Lambda$
      &
      & 
      & $2 \BeNul \!+\! \frac{5}{3}\BeDe$
      & $\BeDe$
      & 0
      & 0                                         \\[\MySep]
    $ K \Sigma$
      &
      &
      &
      & $2 \BeNul \!+\! \BeDe \!-\! 2 \BeFe$
      & 0
      & 0                                         \\[\MySep]
  \hline
    \rule{0em}{2.75\MySep}
          $ \pi N$ 
          \rule{0em}{2.75\MySep}
      & 
      &
      &
      &
      & $2 \BeNul \!+\! \BeDe \!+\! \BeFe$
      & $\BeDe - \BeFe$                           \\[\MySep]
    $ K \Sigma$
      &
      &
      &
      &
      &
      & $2 \BeNul \!+\! \BeDe \!+\! \BeFe$
  \end{tabular}
\end{table}

\begin{table}[htb]
  \centering
  \caption{The $C_{i j}^{(\chi {\rm b})}$ coefficients for the $S=0$, $I=1/2$ and $I=3/2$ channels, 
      $C_{j i}^{(\chi {\rm b})} \!= C_{i j}^{(\chi {\rm b})}$. }
  \vspace{0.75cm}

\begin{tabular}{c|cccc|cc}
    & $\pi N$                                           
    & $\eta n$                                          
    & $K\Lambda$                                        
    & $K\Sigma$                                         
    & $\pi N$                                           
    & $K\Sigma$                              \\[\MySep] 
  \hline
    \rule{0em}{2.75\MySep}
          $\pi N$
          \rule{0em}{2.75\MySep}                        
      & 0 %
      & $\frac{4}{3} (\BeDe + \BeFe)$
      & 0
      & 0
      & 0
      & 0                                         \\[\MySep]
    $ \eta n$
      &  
      & $\frac{8}{9} (\BeDe - 3\BeFe)$
      & $\frac{4}{9} (\BeDe + 3\BeFe)$
      & $- \frac{4}{3} (\BeDe - \BeFe)$
      & 0
      & 0                                         \\[\MySep]
    $ K \Lambda$
      &
      & 
      & 0
      & 0
      & 0
      & 0                                         \\[\MySep]
    $ K \Sigma$
      &
      &
      &
      & 0
      & 0
      & 0                                         \\[\MySep]
  \hline
    \rule{0em}{2.75\MySep}
          $ \pi N$ 
          \rule{0em}{2.75\MySep}
      & 
      &
      &
      &
      & 0
      & 0                                         \\[\MySep]
    $ K \Sigma$
      &
      &
      &
      &
      &
      & 0
  \end{tabular}
\end{table}

\begin{table}[htb]
  \centering
  \caption{The $C_{i j}^{({\rm GOv})}$ coefficients for the $S=0$, $I=1/2$ and $I=3/2$ channels, 
       $C_{j i}^{({\rm GOv})} \!= C_{i j}^{({\rm GOv})}$.}
  \vspace{1em}

\begin{tabular}{c|cccc|cc}
    & $\pi N$                                           
    & $\eta n$                                          
    & $K\Lambda$                                        
    & $K\Sigma$                                         
    & $\pi N$                                           
    & $K\Sigma$                              \\[\MySep] 
  \hline
    \rule{0em}{2.75\MySep}
          $\pi N$
          \rule{0em}{2.75\MySep}                        
      & 0 %
      & $\BeDe + \BeFe$
      & 0
      & 0
      & 0
      & 0                                         \\[\MySep]
    $ \eta N$
      &  
      & $- 4\BeNul \!-\! \frac{10}{3} \BeDe \!+\! 2\BeFe$
      & $- \frac{1}{6} (\BeDe \!+\! 3 \BeFe)$
      & $\frac{1}{2} (\BeDe \!-\! \BeFe)$
      & 0
      & 0                                         \\[\MySep]
    $ K \Lambda$
      &
      & 
      & 0
      & 0
      & 0
      & 0                                         \\[\MySep]
    $ K \Sigma$
      &
      &
      &
      & 0
      & 0
      & 0                                         \\[\MySep]
  \hline
    \rule{0em}{2.75\MySep}
          $ \pi N$ 
          \rule{0em}{2.75\MySep}
      & 
      &
      &
      &
      & 0
      & 0                                         \\[\MySep]
    $ K \Sigma$
      &
      &
      &
      &
      &
      & 0
  \end{tabular}
\end{table}

The exact values of the parameters involved in the tables have to be fixed either 
by relating them to physical observables or in fits to a broader 
set of experimental data. The exact procedure employed in our work 
is described in Section \ref{sec:fits}.



\bibliographystyle{model1a-num-names}
\bibliography{<your-bib-database>}



\end{document}